# ATOMIC TRANSITION PROBABILITIES FOR UV AND OPTICAL LINES OF Gd II[3]

(Short Title: TRANSITION PROBABILITIES OF Gd II)

G. T. Voith[1], E. A. Den Hartog[1], and I. U. Roederer[2]

[1]Department of Physics, University of Wisconsin – Madison, 1150 University Ave, Madison, WI 53706; eadenhar@wisc.edu; gtvoith@wisc.edu

[2]Department of Physics, North Carolina State University, Raleigh, NC 27695; iuroederer@ncsu.edu

ORCIDS:

E. A. Den Hartog:    0000-0001-8582-0910

I. U. Roederer       0000-0001-5107-8930

G. T. Voith          0009-0008-8263-968X

**Abstract**

We report new branching fraction measurements for 156 ultraviolet and optical transitions of Gd II. These transitions range in wavelength (wavenumber) from 2574 to 6766 Å (38838 – 14777 cm$^{-1}$) and originate in one odd-parity and 11 even-parity upper levels. Nine of the 12 levels, accounting for 126 of the 156 transitions, are studied for the first time. Branching fractions are determined for three levels studied previously for the purpose of comparison. The levels studied for the first time are high lying, ranging in energy from 36845 to 40774 cm$^{-1}$. The branching fractions are determined from emission spectra from two different high-resolution spectrometers. These are combined with radiative lifetimes reported in an earlier study to produce a set of transition probabilities and log(*gf*) values with accuracy ranging from 5% to 30%. Comparison is made to experimental and theoretical transition probabilities from the literature where such data exist. Abundances derived from these new log(*gf*) values for 21 Gd II lines in two metal-poor stars yield results consistent with previous studies, and they demonstrate that the new log(*gf*) values can be used in stellar abundance analysis as a self-consistent extension of previous work.

1. **Introduction**

Stars are known to be the builders of the elements in our Universe. The first stars in the universe formed from the hydrogen, helium, and trace of lithium produced by the Big Bang. During their lifetimes and subsequent supernova explosions, they synthesized heavier elements and returned those "metals" to space. The second and later generations of stars, referred to collectively as metal-poor stars, formed from the ashes of the first stars. Studying the abundances of elements in metal-poor stars can provide insight into the elemental formation processes that occurred during the first generations of stars. Elements with atomic number $Z > 30$ can mostly be accounted for by the neutron-capture processes, namely the slow process ($s$-process) and rapid process ($r$-process) (Sneden et.al 2008.) The $s$-process, occurring in low- and intermediate-mass asymptotic giant branch stars (e.g., Karakas 2010), and rapidly rotating massive metal-poor stars (Frischknecht 2016), is reasonably well understood and can be successfully modeled. The $r$-process, on the other hand, takes place in explosive environments of high neutron flux and is less understood and more difficult to model. The astrophysical sites of the $r$-process have been the subject of ongoing debate for decades. Only recently has $r$-process nucleosynthesis been definitively associated with one site – a merging pair of neutron stars (e.g., Drout et al. 2017, Cowan et al. 2021).

The lanthanide elements ($57 \leq Z \leq 71$) have played a crucial role in furthering our understanding of neutron-capture processes and characterizing their sites. These elements have relatively low first ionization potentials (IP), $5.4 \text{ eV} \leq \text{IP} \leq 6.2 \text{ eV}$, and thus are almost completely ionized in the solar photosphere and in the atmospheres of low metallicity giant stars. Their only detectable spectral features arise from their first ionized species. Some of the lanthanides, such as cerium and neodymium, are very accessible spectroscopically in the atmospheres of cool stars because the first ionization state presents tens or hundreds of absorption lines in the optical and ultraviolet (UV) spectral range ($\approx$ 2000–10000 Å; e.g., Moore et al. 1966). Gadolinium, the subject of the current study, falls within this category. Others, such as europium and ytterbium, have fewer but stronger optical transitions that are readily detectable (e.g., Sneden et al. 2009), even when the elemental abundance is low (e.g., Honda et al. 2004). Lanthanide abundances in old, metal-poor stars also indicate the relative contributions of material produced by the $r$- or $s$-processes in prior generations of stars (e.g., Simmerer et al. 2004), the physics responsible for their production (e.g., Roederer et al. 2023), the chemical evolution that occurred in our Galaxy (e.g., Magrini et al. 2018), the galactic components that assembled our Galaxy (e.g., Gull et al. 2021), and the heat budgets of exoplanets (e.g., Wang et al. 2020). Lanthanides also dominate the late-time opacity of the ejecta from merging neutron stars (e.g., Kasen et al. 2013; Tanvir et al. 2017) and potentially even some absorption features (Rahmouni et al. 2024; Vieira et al. 2024), providing the key evidence that directly links $r$-process nucleosynthesis to this site.

Abundance analyses that are accurate to approximately 0.1 dex are sufficient for most applications in stellar astrophysics. Greater accuracy is rarely attainable because of systematic effects from the stellar atmosphere models. Many factors must come together to yield accurate abundances, but among them is the need for accurate transition probabilities for lanthanide elements across a wide range of wavelengths and excitation potential. A wide range of wavelengths is helpful when conducting chemical abundance analyses of stars, because astronomical surveys often span different, and sometimes narrow (~ 100 Å), wavelength ranges. These surveys benefit from the existence of lists of lines suitable for abundance analysis that fall in the predetermined survey wavelength ranges, which are rarely designed with particular heavy-element transitions in mind. Having lines available over a range of excitation potential can be a useful diagnostic. If the line-by-line abundances change systematically with excitation potential that may be indicative of non-LTE effects or some other problem with the stellar atmospheric model.

The most commonly used modern approach to determine experimental transition probabilities (Einstein *A*-values) is to combine an upper level's branching fractions and radiative lifetime. Branching fractions (BFs) are the normalized relative intensities of an upper level's transitions. These relative intensities are determined through high-resolution spectra and the radiative lifetimes are determined through time-resolved laser-induced fluorescence (TRLIF). The current study uses this methodology to determine *A*-values and log(*gf*)s for singly-ionized Gadolinium (Gd, $Z$=64.) In studies of stellar lanthanide abundances, often only the strongest transitions can be observed in the blended spectrum of a star. Even so, determining accurate *A*-values for these strong transitions requires the correct BF normalization, which in turn requires accurate and complete measurements for all branches from an upper level, including the weaker branches. In this study, emphasis is placed on acquiring high S/N spectra in order to accomplish this task.

Previous experimental work on Gd II includes several papers determining radiative lifetimes, and two publications on measured transition probabilities employing modern methods. Of these, the work of Den Hartog et al. (2006; hereafter DH06) is the most complete in regard to both lifetimes and transition probabilities. This was an earlier study from our University of Wisconsin – Madison (UW) group. A TRLIF technique was used to measure the radiative lifetimes for 49 even-parity levels and 14 odd-parity levels. These were combined with branching fractions measured using Fourier transform spectroscopy to determine transition probabilities for 611 lines of Gd II. The most recent measurement of branching fractions (Wang et al. 2014) determined transition probabilities and oscillator strengths (log(*gf*)s) for 12 levels in Gd II using a combination of lifetimes reported in DH06, Wang et al. (2013) and Feng et al. (2011). The Gd II levels studied in Wang et al. (2014) contained just one level with lifetime and BFs previously studied in DH06. The remaining 11 were new BFs measurements and used lifetimes from Wang et al. (2013) or Feng et al. (2011) that were not studied in DH06. Each of these two lifetime studies used a TRLIF technique and agreed well with DH06 for each of five total re-measured levels. In a thesis by Ryder (2011), laser-induced breakdown spectroscopy was used to determine transition probabilities of Gd II and compared them to DH06; a general agreement within uncertainty was concluded.

In addition to the experimental work done, Radžiūtė et al. (2020) calculated energy levels and transition probabilities for some singly-ionized rare earths ($59 \leqslant Z \leqslant 64$) using multiconfiguration Dirac-Hartree-Fock and relativistic configuration-interaction methods. In their paper, they compared their *A*-values for Gd II to those produced in DH06 and found good agreement for strong transitions. A similar comparison is made in this study in Section 3.

Of the 63 total levels that have radiative lifetime measurements in DH06, BF measurements from 52 levels were completed. The 11 without BF measurements had either a strong UV branch beyond the limit of their Fourier transform spectrometer (FTS) spectra or had a severely blended branch. In 2012, the UW 3m High-Resolution Echelle Spectrograph (ECH) was introduced, which has excellent UV capability. This spectrograph has allowed our group to determine BFs for nine new levels that were previously out of range in DH06. The spectra obtained with the ECH are used alongside 15 of the original 16 FTS spectra from DH06.

In Section 2 below, we describe our BF measurements, the two spectrometers used, calibration techniques, and blend analysis. We present results and make comparisons to literature in Section 3. In Section 4 we apply the new data to identify lines that may be useful as Gd II abundance indicators in metal-poor stars. In Section 5 we summarize our conclusions.

## 2. Branching Fractions of Gd II

A BF for a transition between an upper level *u* to a lower level *l* is calculated by taking the ratio of the *A*-value of that transition, $A_{ul}$, over the sum of *A*-values for all transitions from that upper level. This is equivalent to the ratio of the calibrated line intensity of that transition, $I_{ul}$, over the sum of the calibrated intensities of every transition from that upper level:

$$BF_{ul} = \frac{A_{ul}}{\sum_l A_{ul}} = \frac{I_{ul}}{\sum_l I_{ul}} \qquad (1)$$

where $I_{ul}$ is in units proportional to photon counts s$^{-1}$. By definition, the BFs associated with *u* sum to 1. Relative intensities where the sum in the denominator is over fewer than the full set of transitions are referred to as branching ratios (BRs.) The radiative lifetime of the upper level, $\tau_u$, is equivalent to the inverse sum of all transition *A*-values attached to the upper level:

$$\tau_u = \frac{1}{\sum_l A_{ul}} \qquad (2)$$

The lifetimes for the levels used in this study were measured previously by DH06. These are used to determine final *A*-values and log(*gf*)s from the BFs by substituting eq. 2 into eq. 1 and following the relations in Martin et al. (2023):

$$A_{ul} = \frac{BF_{ul}}{\tau_u} \quad ; \quad \log(gf) = \log\left(\frac{1.4492 g_u A_{ul}}{\sigma^2}\right) \qquad (3)$$

where $A_{ul}$ is the transition probability in s$^{-1}$, $\tau_u$ is the radiative lifetime of the upper level in seconds, $g_u$ is the degeneracy of the upper level, and $\sigma$ is the transition wavenumber in cm$^{-1}$.

The energy level structure of $Gd^+$ is sufficiently well known[4] and there are few missing low-lying levels of either parity that might give rise to missing branches in the current study. All possible dipole-allowed transitions obeying the $\Delta J$ and parity change selection rules are investigated, and all observed transitions are analyzed. Some weak transitions that have >30% final uncertainty are not included in the tables but are included in the BF normalization. The sum of these "residual" transitions ranges between 0% and 8% for the levels in this study.

### *2.1 Two Spectrometers*

This study used spectra from two different high-resolution spectrometers: archived spectra with wavelength ranging from ~290 nm in the near-UV through 6 μm in the infrared (IR) collected on the 1 m FTS at the National Solar Observatory (NSO) in Kitt Peak, AZ [5] (Table 1); and new spectra with wavelength range 235 nm – 436 nm collected on the UW 3 m Echelle Spectrograph (ECH) (Table 2). Together, these spectra cover the entirety of the wavelength range required for the study of transitions from these high-lying energy levels of Gd II. FTSs in general are an excellent option for measuring BFs as they offer exceptional wavenumber accuracy at high resolution. The 1 m FTS at NSO, in particular, has wavenumber accuracy of **about** 1 part in $10^8$, a limit of resolution as low as 0.01 cm$^{-1}$, the capability of recording a 1 million part spectrum in 10 minutes (Brault 1976), and a broad spectral range. The FTS spectra listed in Table 1 are a subset of those used in the earlier DH06 study. Here we use the first 15 of the 16 spectra listed in Table 2 of that paper.

This study used a combination of spectral sources. In the archived FTS spectra listed in Table 1, three types of sources were used: commercial hollow cathode discharge lamps (HCL), a custom water-cooled HCL, and an electrodeless discharge lamp (EDL). The commercial HCL was air-cooled and was run at currents no higher than 30 mA, while the custom water-cooled HCL was operated at much higher currents, between 290 mA and 300 mA. These high current spectra as well as those from the EDL have superior S/N, particularly on weak lines, compared to those from the commercial HCLs. They are useful in determining BFs for lines that would otherwise be lost in the noise. The spectra taken on the ECH (Table 2) were taken solely using air-cooled commercial HCLs run at a range of currents no higher than 25 mA. Depending on lamp design, lines from various other species appear in the spectra alongside the ionizations of Gd. Specifically, Ar, Ne, Fe, U, and $I_3$ are discussed further in section 2.3, which details blend analysis.

Quantum statistical (Poisson) noise arises as the square root of a signal's strength. For the FTS, which is an interferometric device, the Poisson noise from all lines spreads out evenly across the Fourier-transformed spectrum. This is known as multiplex noise and can become a problem at low source current because the noise from strong branches can then overshadow the weaker branches. The step often taken to compensate for this is to increase the source current until the weaker lines have a higher signal than the multiplex noise. However, self-absorption becomes an issue at higher source currents. At high currents, there are more sputtered ions in the lamp. Photons released from strong transitions connected to near-ground and metastable lower levels have an increased chance to become reabsorbed by other ions of the same species before leaving the lamp. Evidence of self-absorption appears when comparing the BR of a strong branch across high-current spectra to its BR across low-current spectra. A strong branch, if it is self-absorbed, will have lower BR in high-current spectra than in low-current spectra relative to the weaker branches. When self-absorption is observed, the high-current measurements of the strong branch are excluded from the data set.

---

[4] Throughout this article and associated tables, we use the energy levels from Martin et al. (1978) downloaded from the National Institute of Standards and Technology Atomic Spectra Database (NIST ASD; Kramida et al. 2024, available at **https://physics.nist.gov/asd**). Transition wavenumbers are calculated from the difference in level energies, and these are converted to air wavelengths using the five parameter formula for the standard index of air (Peck & Reeder 1972).
[5] This instrument was decommissioned in 2012, but all spectra recorded with it are archived and publicly available at https://nispdata.nso.edu/ftp/FTS_cdrom/.

The ECH utilizes a McPherson model 2173 3 m focal length Czerny-Turner spectrometer. The standard grating is replaced with a large, coarse echelle grating blazed at high angle (128 mm x 254 mm ruled area, 23.2 grooves mm$^{-1}$ and blaze angle[6] of 63.5º). The entrance slit assembly is modified to hold a precision entrance pinhole. Typical operation uses a 50 μm pinhole. On its exit port is a custom 0.5 m focal length prismatic order separator. The optical plane of the order separator is orthogonal to the optical plane of the main instrument. This 90º rotation results in the cancellation of spherical aberration, the primary aberration of the 3 m spectrometer, by that of the order separator. Combined, these disperse the light into a two-dimensional spectrum imaged onto a square 2048 x 2048 pixel UV-sensitive CCD array with the high dispersion direction running in one direction and successive orders arrayed side-by-side in the other direction. Further discussion of the instrument can be found in Wood & Lawler (2012). A significant advantage of the ECH is its sensitivity in the UV. As a dispersive instrument, it does not suffer from multiplex noise, allowing the measurement of weak lines with good S/N even at low source currents. This characteristic makes it an excellent tool for avoiding self-absorption errors that can occur in high-current spectra.

In addition to its high UV sensitivity, the ECH is also characterized by high dispersion and a high resolving power of ~250,000. This high resolving power is very advantageous when studying a complicated and crowded spectrum such as Gd II. However, the high dispersion results in the blaze envelope of the ECH grating being too large to be imaged on a single CCD frame. It is therefore required to take spectra at three grating positions to capture the full blaze envelope. The grating settings are chosen such that, when combined, all three frames cover the entire span of each of the orders in the high-dispersion direction while also allowing for overlap between successive frames. Matching lines between overlapping frames allows for relative scaling of the frame intensities. This scaling factor is typically near unity and determined with relatively small uncertainty (~1% – 5%). In addition to the three frames, one frame (both HCL and D$_2$) is repeated for redundancy to account for any possible lamp drift that may have occurred, for a total of four CCD frames per spectrum.

*2.2 Calibration*

---

[6] The blaze angle of an echelle grating is the angle between the plane of the grating and the facet.

A feature of using two differently designed instruments with independent radiometric calibrations is that the comparison between BRs measured on both instruments helps assess whether the systematic uncertainty assigned to each is realistic. The FTS spectra in Table 1 were calibrated for the DH06 study using a combination of methods for redundancy, which are listed in the final column of the table. The same radiometric calibrations for the DH06 study were used again in this study. For deeper discussion of the FTS calibration methods, we refer the reader to DH06. The Ar I and II method uses sets of well-known neutral and singly ionized argon BRs established by Adams & Whaling (1981), Danzmann & Kock (1982), Hashiguchi & Hasikuni (1985), and Whaling et al. (1993). Comparing these known BRs of lines ranging over 4300 – 35000 cm$^{-1}$ to their measured intensities provides a calibration curve for each spectrum. The advantage of this method is that it accounts for wavelength-dependent factors of all components in the light path and reflections that may affect signal detection. In the spectra recorded prior to 2000, a tungsten strip lamp spectrum calibrated as a spectral radiance (W m$^{-2}$ sr$^{-1}$ nm$^{-1}$) standard was recorded either shortly before or shortly after every Gd II spectrum. Similarly done in the 2002 spectrum, a tungsten-quartz-halogen lamp was calibrated as a spectral irradiance (W m$^{-2}$ sr$^{-1}$ nm$^{-1}$ at a specified distance) standard. The Ar I and II method is superior in the UV as the tungsten lamps are neither hot nor bright enough to be useful in the UV. However, they do work as useful redundancies in the visible and IR range. The top panel of Figure 1 shows an example of these two calibration methods for spectrum 6 listed in Table 1. The continuum lamps are particularly helpful in spectral regions where the instrument sensitivity is changing rapidly as a function of wavenumber, which can be difficult to capture with the Ar I and II BRs. In the current study, we use the sensitivity curves as determined by DH06.

The ECH calibration relies on a NIST traceable deuterium (D$_2$) lamp continuum. Immediately after each Gd II frame, a deuterium spectrum is taken, with the only change between the HCL and D$_2$ spectra being the rotation of a flat mirror on a kinematic mount. In previous studies done with the ECH, the regularly used D$_2$ lamp was calibrated against a rarely used D$_2$ lamp. In recent years, a shift has been made to using a detector-based calibration technique. The technique, explained in detail in Den Hartog et al. (2023), utilizes a NIST-calibrated photodiode. The calibration of the photodiode is inherently more stable than that of a continuum lamp which undergoes changes due to aging and UV damage over time.

The software used to analyze both the FTS and ECH spectra is much the same as that used in DH06; however, the ECH required some modification to account for the two-dimensionality of its spectrum. As in DH06, it takes in a list of all known levels of Gd II and, while accounting for parity and $\Delta J$ selection rules, calculates all possible transitions from the level under study to lower levels. The user can call up any indexed transition and view a small portion surrounding the transition wavenumber **in** each corresponding spectrum. It then allows for simple numerical integration of the line with user-defined limits. The difference is that when an indexed transition is called, the ECH program produces a two-dimensional plot of the spectrum centered around the transition wavenumber. In this perspective, the user can cut away rows of pixels along the high dispersion direction to isolate the correct order. When the order is isolated, the two-dimensional perspective plot is summed along the low dispersion direction. This produces a 1D plot where the user can define limits along the high dispersion direction and integrate as one would in the FTS. Immediately following this integration, the program shifts to the D$_2$ spectrum and locates the same position where the integrated Gd II line was. The user then integrates across the width of the order in the low-dispersion direction and 25 pixels in the high-dispersion direction along the order in question. Note that for both the line spectrum and continuum spectrum, the integration captures intensity from only one order in the low dispersion direction. The reason for the fixed 25 pixels in the high-dispersion direction for the D$_2$ integration is to ensure consistency of the D$_2$ integral for all lines from a given upper level. The relative sensitivity of the instrument at a given wavelength in a given order is given by

$$sensitivity \propto \frac{I_{D2} \times \sigma^2}{IRR_{D2}} \qquad (4)$$

where $I_{D2}$ and $IRR_{D2}$ are the integrated intensity and calibrated irradiance of the D$_2$ lamp, respectively. One of the factors of $\sigma$ in equation 4 is needed in the numerator to convert the IRR from Watts to photons s$^{-1}$ and the other to remove the proportionality to wavelength of the fixed 25 pixel integral, $I_{D2}$. The bottom panel of Figure 1 shows the measured sensitivity at the peak of the grating response in selected orders, showing the change in sensitivity in the low dispersion direction. The rapidly changing sinc$^2(\Delta\lambda)$ behavior in the high-dispersion direction is not depicted, but one could imagine the full sensitivity curve as being saddle-shaped, with the curve shown in Figure 1 being the centerline profile. The Gd II line intensity is divided by the relative sensitivity. Similarly done in the FTS where the line integral is divided by the spectral sensitivity from the calibration curve generated by the Ar I and II integrations. These determine the BRs for each transition on each individual spectrum.

The uncertainty of a BR is determined from the inverse of the S/Ns, the standard deviation of the weighted mean and the systematic uncertainty of the calibration. The latter is conservatively estimated to be 0.001% per cm$^{-1}$ between the line of interest and the dominant line(s) from the upper level. In the case where there is a single dominant line, the cumulative systematic uncertainty is attributed entirely to the weak line(s). Where there are multiple strong branches, they share this uncertainty between them in proportion to the inverse of the BFs. For example, two equally strong lines separated by 6000 cm$^{-1}$ would each have 3% systematic uncertainty ascribed to them. However, two lines with 0.33 and 0.67 BRs separated by 6000 cm$^{-1}$ would have 4% and 2% systematic uncertainty, respectively.

Branches in the overlap wavelength range from ~290 nm to 436 nm are used to put the ECH and FTS BRs on the same scale. The complete set of BRs are then renormalized to yield BFs. The BF uncertainty of the ECH-only lines is just the BR uncertainty, while that of the lines in the overlap and the FTS-only lines have an added rescaling uncertainty added in quadrature with the BR uncertainty. For cases where there is a significant far-UV branch omitted from the FTS analysis, FTS-only lines have an additional systematic uncertainty added in quadrature.

## 2.3. Blends

Gadolinium, as with all lanthanides, has a complex spectrum. This complexity arises from its open *s, p, d,* and *f* shells which lead to many low-lying configurations of opposite parity. The resulting high density of lines increases the possibility of blended lines in the spectrum. To measure accurate branching fractions, these blends must be addressed. A blend separation technique, based on a least-squares fit analysis, was implemented in the current study following the methods outlined in Den Hartog et al. (2019). It relies on the observation that an upper level's population varies uniquely with respect to source conditions. This is reflected in changes to the relative intensity of the blended line versus blending partner as source conditions differ **in** the various spectra. Figure **2** illustrates this concept with a small portion of five echelle spectra taken under different source current and buffer gas combinations. The five Gd-Ar and Gd-Ne spectra shown are all normalized at the 3757.74 Å line originating from a high-lying Gd II upper level (38922 cm$^{-1}$) making clear the relative intensity variation of the nearby Gd I line at 3757.94 Å and a Gd II line at 3758.31 Å originating from a lower-lying upper level (30027 cm$^{-1}$).

In our analysis, every possible transition, according to parity and $\Delta J$ selection rules, is calculated, and the position is marked on the displayed spectrum during analysis. When a potential blend is noted in the spectrum, it is possible to look up the blending partner's upper level and indexed transitions. BRs of clean/unblended branches are measured from the upper levels associated with both the line of interest and the blending partner. The clean BRs are then compared directly to that of the blend across all spectra. The greatest range of intensities are seen between Gd/Ar and Gd/Ne spectra, which is why the ECH, having a calibrated set of Gd/Ne spectra, is used for the majority of blend separations. One clean branch from each upper level is enough to generate an exact solution for a blending fraction for each spectrum. In cases where multiple clean branches are available, the blend can be analyzed using a least-squares analysis. It is possible that the upper level associated with the blending partner has no clean lines to use in the comparison. In this case, it has shown equal promise to use a near-lying upper level from the same multiplet as the one in question, as their BRs have been observed to vary similarly over source conditions (Den Hartog et al. 2019).

Blends with other contaminant species can often be resolved by a change of source. The buffer gas used in each lamp produces either Ar or Ne lines to appear in the spectra. If there is an Ar blend, then only the Gd/Ne spectra are used for that branch. There are Gd/Ne spectra listed in both Table 1 and Table 2; however, the ECH has the only calibrated Gd/Ne spectra out of the two spectrometers. This is due to the nature of the $D_2$ calibration used in the ECH, as opposed to the Ar I and II calibration used in the FTS. The Gd/Ne spectra taken on the FTS were used strictly as comparisons for such blends. These buffer gas lines were observed to originate from neutral and singly-ionized Ar and Ne. Of the levels analyzed in this study, there were no instances where both buffer gases had blended with the same branch.

Fe also appears as a blending candidate in some of our spectra. Some commercial HCLs of lanthanide elements, such as the ones used in this study, have a cathode constructed of Fe with a thin inner lining of the lanthanide element, in this case Gd. This can result in some Fe I-II lines appearing in the spectra, particularly after the lamps are heavily used. This is why neutral and ionized Fe were considered as possible contaminants in this study. That being said, very few Fe lines were observed – only a few strongest lines of Fe I appeared weakly in some spectra. In the FTS spectra, the custom water-cooled HCL had no Fe in the cathode, and thus, there was no Fe observed in those spectra with indices 6-11. Spectrum 15 was taken on an EDL with a Gd/$I_3$ sample and had no observed Fe influence, while spectrum 14, which was taken on an EDL with a Gd/U sample, did have observed Fe I lines. Any traces of triiodide and uranium were dealt with in similar fashion to the buffer gas blends. The ECH spectra were taken solely using commercial HCLs.

## 3. Results

The BFs determined in the current study are reported in Table 3. Twelve upper energy levels were studied, with one out of the twelve being odd parity. Three of the even-parity levels were studied previously by DH06 and are included in this study to check for consistency of analysis. The three repeated levels are those at 30366.818, 30996.851, and 34178.776 cm$^{-1}$. The measurements are, of course, not entirely independent as they use the same FTS spectra with the same radiometric calibration as determined for the DH06 study. However, the new BFs are determined with a reanalysis of the FTS data by different personnel as well as using data from the ECH. The reanalysis consisted of a weighted average between remeasured BRs from 15 of the original 16 FTS spectra from DH06 and the BRs from the ECH spectra. The comparison is shown in Figure **3**. For the majority of these branches, there is good agreement. A few outliers, all of which have BFs less than ~0.03, do not agree within the combined uncertainties. These weak lines have better S/N and lower uncertainty in the ECH measurements, and we recommend the new measurements accordingly. We use the radiative lifetimes of DH06 to determine *A*-values and log(*gf*)s from our BFs according to Equation 3. The uncertainty of the *A*-value is the uncertainty of the BF and that of the lifetime added in quadrature. These are presented in Table 4 organized by increasing wavelength in air.

Radžiūtė et al. (2020) have done theoretical calculations of Gd II transition probabilities and compared them to those that overlapped with DH06. In their comparison, they noted an underestimation of many of the weaker branches. We now compare the new transition probabilities to those calculated in Radžiūtė et al. Comparison was done by matching the electron configurations and term for every transition in this study to those in the theory. Transition probabilities measured in this study were plotted against those from matching configurations in Radžiūtė et al. Not all level configurations in this study could be matched to those in Radžiūtė et al. A comparison of those that could be matched is shown in Figure **4**. The solid line in this figure represents perfect agreement while the dotted lines represent plus and minus a factor-of-two. There is satisfactory agreement for most of the *A*-values for strong lines ($A > 1E^7$). The agreement is less satisfactory for weaker transitions, which are difficult to calculate accurately because of the high degree of configuration mixing.

## 4. Gd Abundance Determination in Two Metal-Poor Stars

We apply these new transition probabilities to high-resolution and high-S/N optical and ultraviolet (UV) spectra of two metal-poor, *r*-process-enhanced stars, HD 222925 and 2MASS J22132050−5137385 (hereafter J2213−5137). The *r*-process-enhanced nature of these stars gives rise to spectra that are rich in lanthanide features, including dozens of Gd II absorption lines. Both stars have been extensively studied previously. HD 222925 exhibits an *r*-process enhancement, characterized by the [Eu/Fe][7] ratio, of +1.32 ± 0.08 (Roederer et al. 2022), and J2213−5137 exhibits [Eu/Fe] = +2.45 ± 0.08 (Roederer et al. 2024). For HD 222925, Roederer et al. (2018, 2022) derived an effective temperature ($T_{\rm eff}$) of 5636 ± 103 K, log of surface gravity (log $g$) of 2.54 ± 0.17 (in cgs units), microturbulent velocity parameter ($v_t$) of 2.20 ± 0.20 km s$^{-1}$, and metallicity ([Fe/H]) of −1.46 ± 0.10. For J2213−5137, Roederer et al. (2024) derived $T_{\rm eff}$ of 5509 ± 72 K, log $g$ of 2.28 ± 0.06, $v_t$ of 2.25 ± 0.10 km s$^{-1}$, and [Fe/H] of −2.20 ± 0.12. We adopt these model atmospheres which we previously interpolated (Roederer et al. 2022, 2024) from the ATLAS9 grid (Castelli & Kurucz 2004).

We revisit the spectra collected by previous studies. Full details of these spectra may be found in the references cited above, and here we provide only a brief summary of key characteristics. Optical spectra were collected using the Magellan Inamori Kyocera Echelle (MIKE) spectrograph (Bernstein et al. 2003) at the Magellan II (Clay) Telescope at Las Campanas Observatory (HD 222925: 3330 ≤ $\lambda$ ≤ 5000 Å at $R \equiv \lambda/\Delta\lambda$ = 68,000 and 5000 ≤ $\lambda$ ≤ 9410 Å at $R$ = 61,000; J2213−5137: 3260 ≤ $\lambda$ ≤ 4780 Å at $R$ = 60,000 and 4780 ≤ $\lambda$ ≤ 8000 Å). S/N ratios near 3500 Å in the MIKE spectra are approximately 80/1 for each pixel in HD 222925 and 70/1 for each pixel in J2213−5137. UV spectra of HD 222925 were collected using the Space Telescope Imaging Spectrograph (STIS; Kimble et al. 1998; Woodgate et al. 1998) on board the Hubble Space Telescope (1936 ≤ $\lambda$ ≤ 3145 Å at $R$ = 114,000). S/N ratios near 2700 Å are approximately 40/1 for each pixel.

Abundances are derived by matching synthetic spectra, generated using a recent version[8] of the MOOG spectrum analysis code (Sneden 1973; Sobeck et al. 2011) and the LINEMAKE linelist generator[9] (Placco et al. 2021), to the observed spectra. The Gd abundances of these stars have been presented previously by the above references, but we have searched the optical and UV spectra of these stars for additional lines presented in Table 4. We check the ≈ 40 lines with relative strength factors, defined as log $\varepsilon$+log($gf$) – $\theta\chi$, where log $\varepsilon$ refers to the log of the number density log$_{10}$ ($N_{\rm Gd}/N_{\rm H}$) + 12.0, $\theta \approx$ 5500/5040 and $\chi$ is the excitation potential of the lower level in eV, > −0.6 or so (DH06). Table 5 presents a list of the Gd II lines we detected in these spectra that are sufficiently free of blends from other species. Six new lines are used to derive the Gd abundance in HD 222925, and five new lines are used for J2213−5137. One of these lines, at 3395.12 Å, is detected in both stars' spectra, and it is shown in Figure **5**. The shaded bands around the best-fit spectrum synthesis curves in Figure **5** indicate variations of ± 0.3 dex, a factor of two. The fits are considerably better than ± 0.3 dex in both cases, and we estimate fitting (statistical) uncertainties of ± 0.15 dex in HD 222925 and ± 0.10 dex in J2213−5137. The uncertainties in the log($gf$) values are considerably smaller than the observational uncertainties, which are typically dominated by our ability to identify the local continuum in the observed spectrum. Roederer et al. (2022, 2024) estimated systematic uncertainties that include the uncertainties in the model atmosphere parameters, which amount to approximately 0.09 dex and 0.12 dex for the mean Gd abundances in HD 222925 and J2213−5137, respectively.

---

[7] Abundance ratios [X/Y] are defined as log$_{10}$($N_X/N_Y$) – log$_{10}$($N_X/N_Y$)$_\odot$, where the $\odot$ indicates the Solar ratio.
[8] https://github.com/alexji/moog17scat
[9] https://github.com/vmplacco/linemake

The new log(*gf*) values yield abundances in agreement with those derived using the DH06 log(*gf*) values, which is unsurprising given the excellent agreement between the BF measurements shown in Figure **3**. For 12 lines in HD 222925, we find log $\varepsilon$ (Gd) = 0.88 ± 0.04 using the DH06 log(*gf*) values and log $\varepsilon$ (Gd) = 0.87 ± 0.04 using the new log(*gf*) values. For 13 lines in J2213−5137, we find log $\varepsilon$ (Gd) = 1.05 ± 0.03 using the DH06 log(*gf*) values and log $\varepsilon$ (Gd) = 1.06 ± 0.03 using the new log(*gf*) values. The log $\varepsilon$ values reported are inverse-square-weighted means, and the uncertainties are the standard error of the mean, which do not include the systematic components of uncertainty discussed above.

The Gd abundances derived from the additional lines covered by the present study are also in excellent agreement with the Gd abundances derived from many more lines studied previously, using atomic data from DH06. For HD 222925, we find log $\varepsilon$ (Gd) = 0.88 ± 0.07 (statistical) based on six new lines, and Roederer et al. (2022) found log $\varepsilon$ (Gd) = 0.82 ± 0.09 (statistical+systematic) based on 44 other optical and UV Gd II lines. For J2213−5137, we find log $\varepsilon$ (Gd) = 1.08 ± 0.06 (statistical) based on five new lines, and Roederer et al. (2024) found log $\varepsilon$ (Gd) = 1.12 ± 0.12 (statistical+systematic) based on 34 other optical Gd II lines.

These results affirm that the new set of log(*gf*) values presented in Table 4 can be used in stellar abundance work as a self-consistent extension of the previous Gd II line list presented by DH06.

## 5. Summary

We report new BF measurements for 156 transitions connected to 12 high-lying levels of Gd$^+$. The BFs are combined with radiative lifetimes from an earlier study to produce a set of *A*-values and log(*gf*)s for those lines with uncertainties ranging from 5 to 30%. Comparison is made to experimental and theoretical data from the literature where available. The new data are employed to determine Gd abundances in two metal-poor stars. Twenty-one lines are found that are strong and unblended enough to yield reliable Gd abundances, and six of these are reported for the first time. The newly determined Gd abundances are in good agreement with previous abundance determinations in those stars using other UV and optical Gd II lines.

## Acknowledgments

E.A.D.H. acknowledges support from the U.S. National Science Foundation (NSF) grant AST 2206050. I.U.R. acknowledges support from grants GO-15657 and GO-17166 from the Space Telescope Science Institute, which is operated by the Association of Universities for Research in Astronomy, Incorporated, under NASA contract NAS5-26555, and from NSF grant AST 2205847.

# Figure Captions

Figure 1: Examples of calibration curves for the FTS (top) and ECH (bottom). The top panel shows redundant FTS sensitivity curves for spectrum 6 from the Ar I & Ar II BR method and from a W strip lamp. The bottom panel shows the sensitivity for spectrum 29 of the ECH as it changes in the low-dispersion direction. The rapidly changing $\text{sinc}^2(\Delta\lambda)$ behavior in the high-dispersion direction is not depicted.

Figure **2**: Samples from several echelle spectra taken under different source conditions and normalized at the 3757.74 Å line of Gd II (12318 - 38922 cm$^{-1}$). The variation of relative intensity of a line at 3758.31 Å originating from a different level of Gd II (3427 - 30027 cm$^{-1}$) and a Gd I line at 3757.94 Å (533 - 27136 cm$^{-1}$) is apparent.

Figure **3**: Logarithmic differences of experimental BFs measured in this study and those from DH06. The solid horizontal line at 0.0 indicates perfect agreement and error bars indicate the uncertainties of the two studies added in quadrature. It should be noted that these studies are not entirely independent. See text for further discussion.

Figure **4**: *A*-values from the theory of Radžiūtė et al. (2020) versus the experimental *A*-values from this study. The solid diagonal line indicates perfect agreement while the dashed lines indicate factor-of-two disagreement.

Figure **5**: Comparison of observed and synthetic spectra around the Gd II line at 3395.12 Å for HD 222925 (top) and J2213−5137 (bottom). The small black squares mark the observed spectrum of each star. The solid red line marks the best-fit abundance, the shaded red band marks variations of ± 0.3 dex from this best-fit abundance, and the black line marks a synthetic spectrum with no Gd.

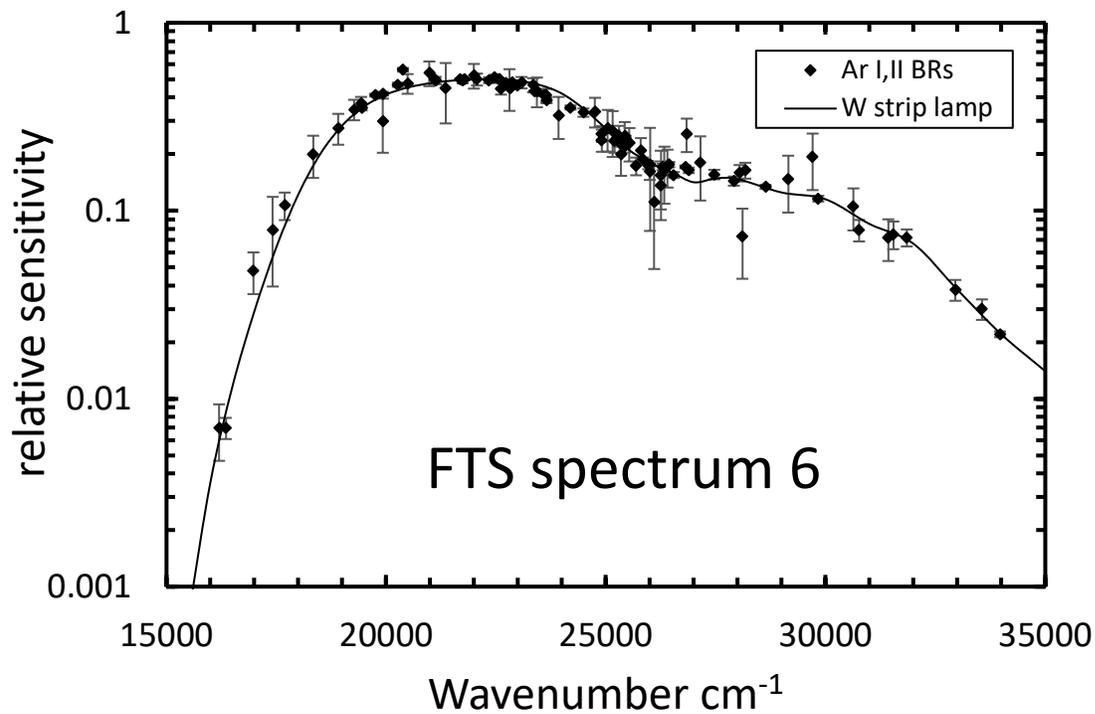

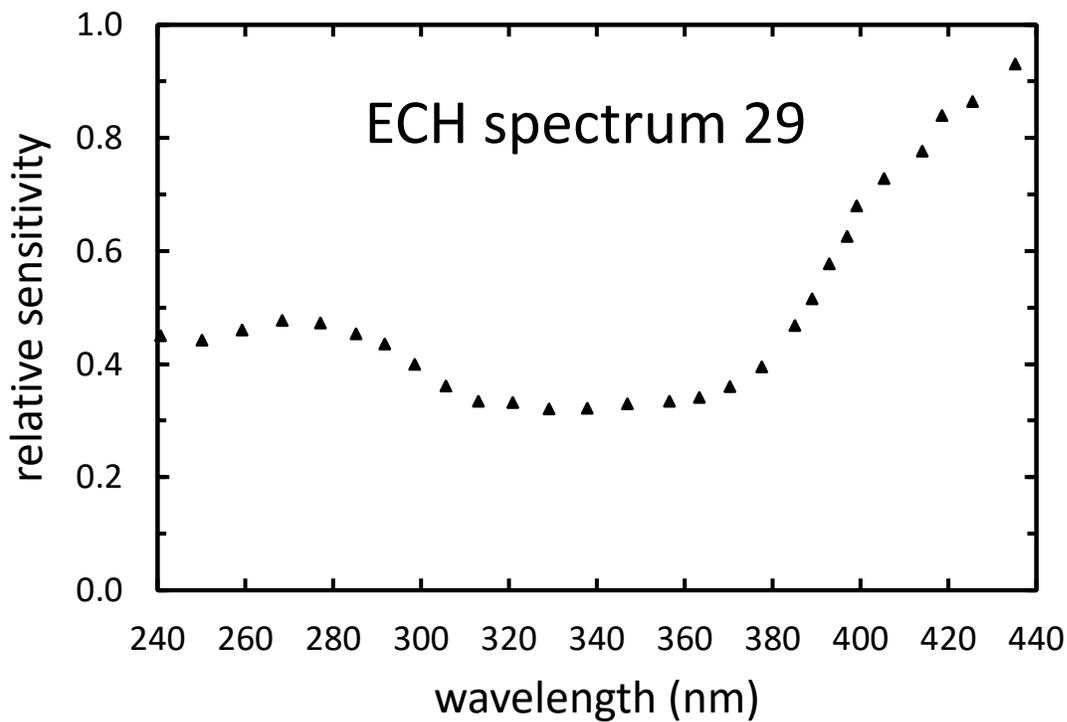

Figure 1: Examples of calibration curves for the FTS (top) and ECH (bottom). The top panel shows redundant FTS sensitivity curves for spectrum 6 from the Ar I & Ar II BR method and from a W strip lamp. The bottom panel shows the sensitivity for spectrum 29 of the ECH as it changes in the low-dispersion direction. The rapidly changing $sinc^2(\Delta\lambda)$ behavior in the high-dispersion direction is not depicted.

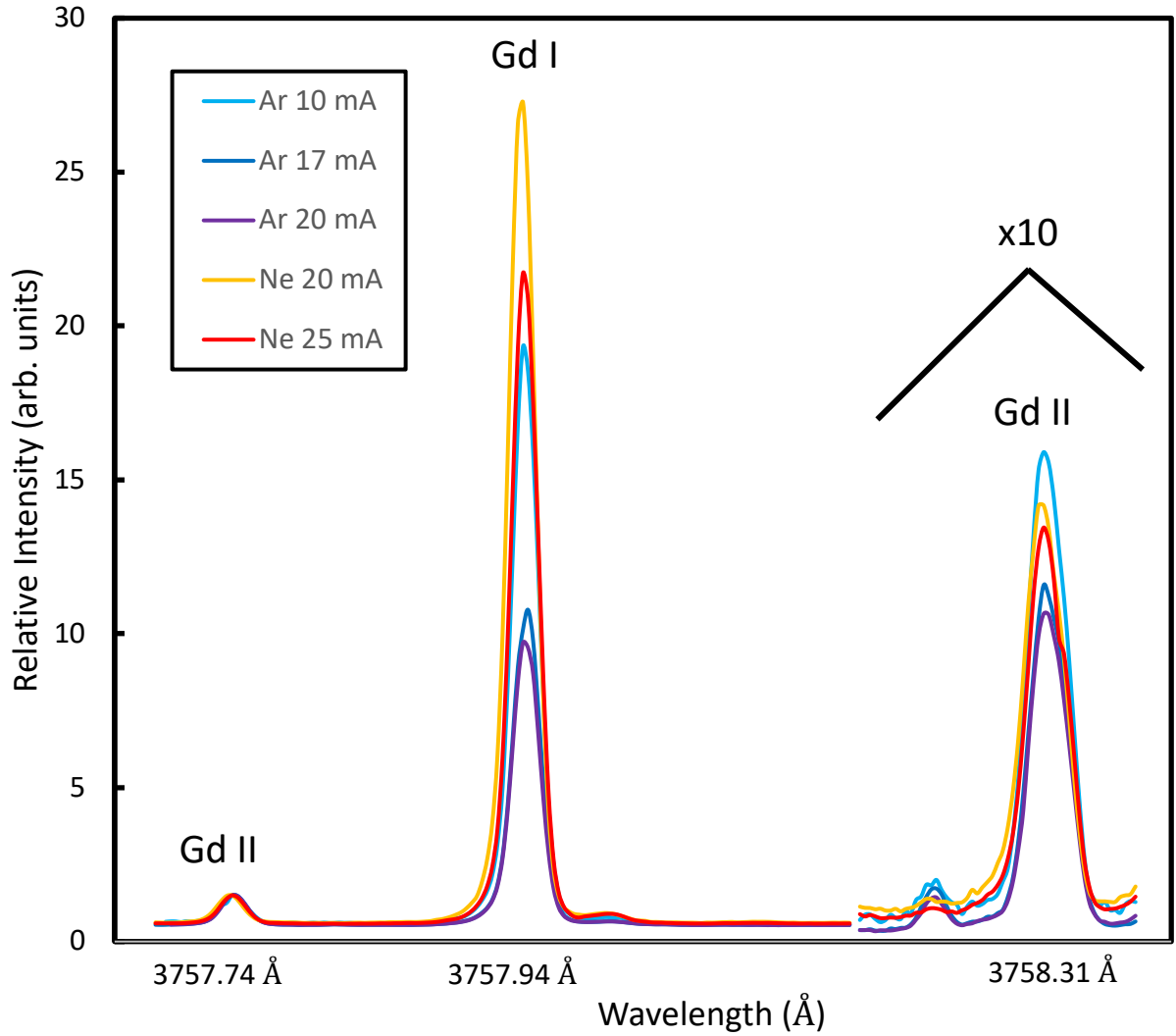

Figure 2: Samples from several echelle spectra taken under different source conditions and normalized at the 3757.74 Å line of Gd II (12318 - 38922 cm$^{-1}$). The variation of relative intensity of a line at 3758.31 Å originating from a different level of Gd II (3427 - 30027 cm$^{-1}$) and a Gd I line at 3757.94 Å (533 - 27136 cm$^{-1}$) is apparent.

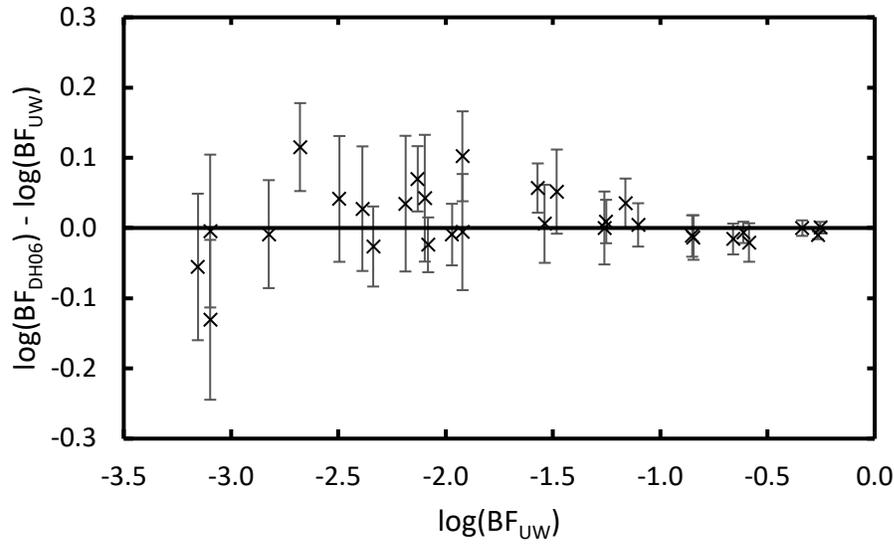

Figure **3**: Logarithmic differences of experimental BFs measured in this study and those from DH06. The solid horizontal line at 0.0 indicates perfect agreement and error bars indicate the uncertainties of the two studies added in quadrature. It should be noted that these studies are not entirely independent. See text for further discussion.

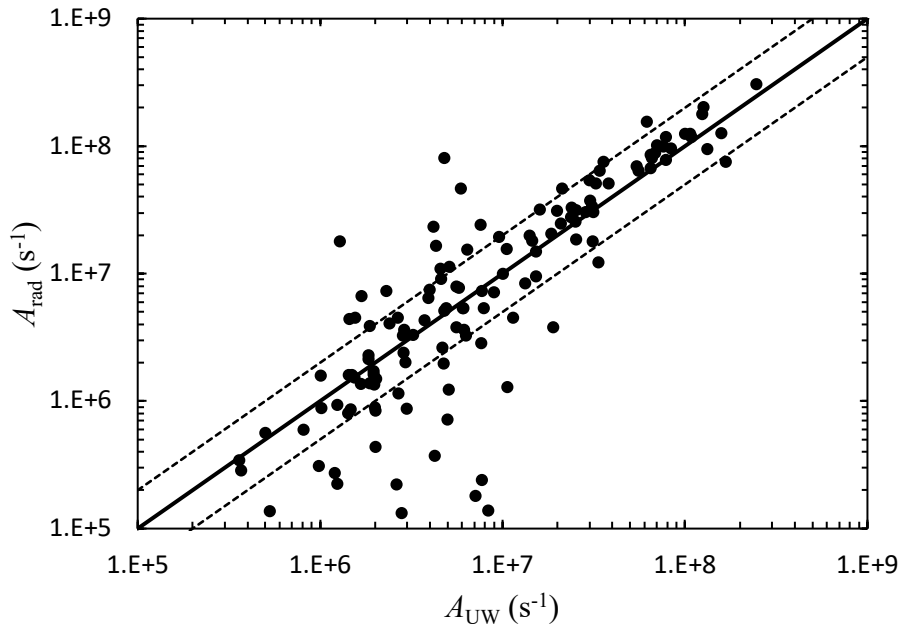

Figure **4**: $A$-values from the theory of Radžiūtė et al. (2020) versus the experimental $A$-values from this study. The solid diagonal line indicates perfect agreement while the dashed lines indicate factor-of-two disagreement.

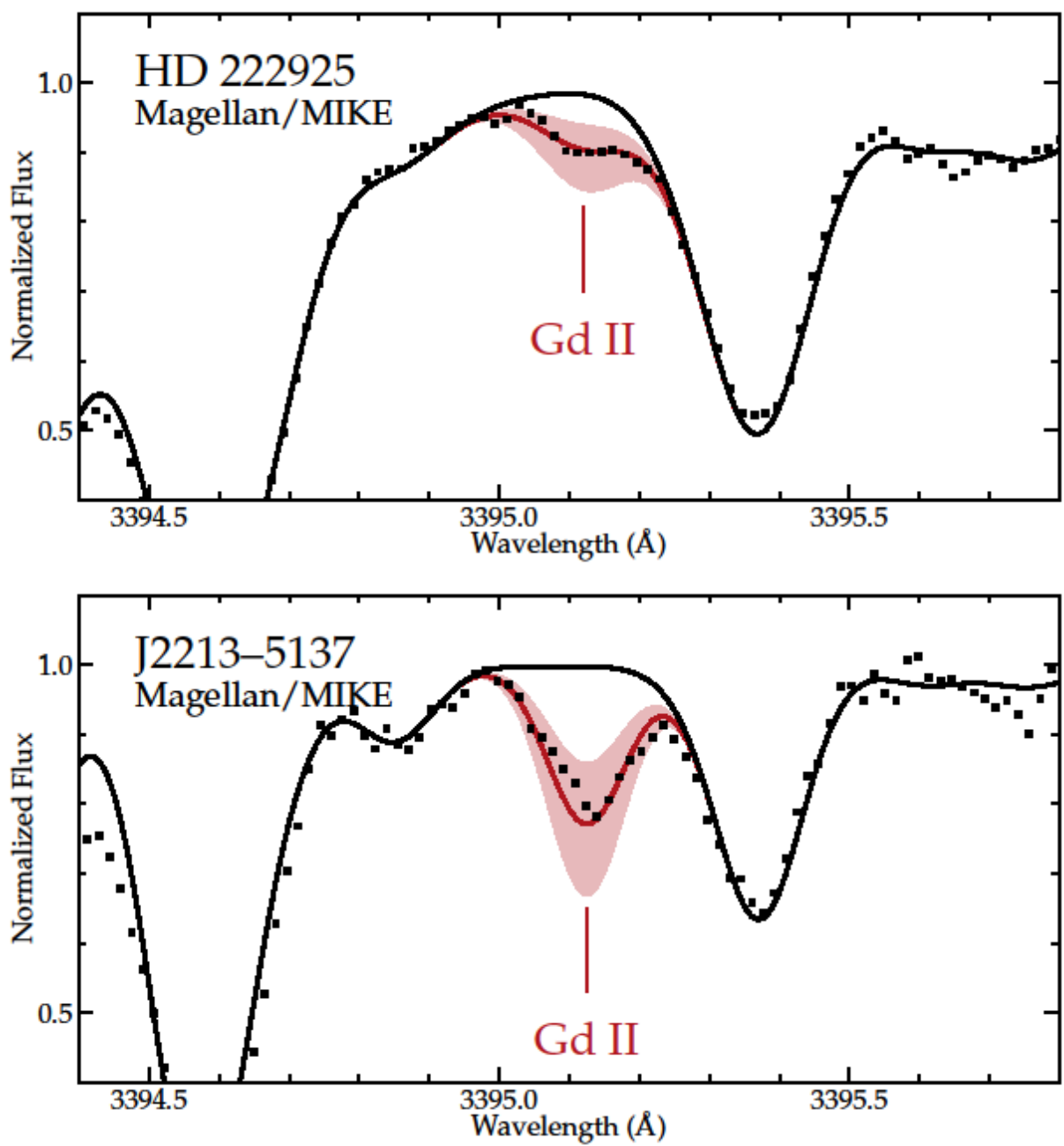

Figure **5**: Comparison of observed and synthetic spectra around the Gd II line at 3395.12 Å for HD 222925 (top) and J2213−5137 (bottom). The small black squares mark the observed spectrum of each star. The solid red line marks the best-fit abundance, the shaded red band marks variations of ± 0.3 dex from this best-fit abundance, and the black line marks a synthetic spectrum with no Gd.

**Table 1**
Fourier Transform Spectra[a]

| Index | Date | Serial Number | Lamp Type[b] | Buffer Gas | Lamp Current (mA) | Wavenumber Range (cm$^{-1}$) | Limit of Resolution (cm$^{-1}$) | # scans | Beam splitter | Filter | Detector[c] | Calibration[d] |
|---|---|---|---|---|---|---|---|---|---|---|---|---|
| 1 | 2002 Feb. 27 | 18 | Commercial HCL | Ar | 26 | 7925-34998 | 0.050 | 50 | UV | | S. B. Si PD | Ar I & II, WQH Lamp |
| 2 | 2000 Feb. 28 | 23 | Commercial HCL | Ar | 23 | 7925-34998 | 0.053 | 16 | UV | | S. B. Si PD | Ar I & II |
| 3 | 2000 Feb. 27 | 18 | Commercial HCL | Ar | 30 | 7925-34998 | 0.053 | 41 | UV | | S. B. Si PD | Ar I & II |
| 4 | 2000 Feb. 27 | 19 | Commercial HCL | Ar | 15 | 7925-34998 | 0.053 | 9 | UV | | S. B. Si PD | Ar I & II |
| 5 | 2000 Feb. 27 | 20 | Commercial HCL | Ar | 12.5 | 7925-34998 | 0.053 | 9 | UV | | S. B. Si PD | Ar I & II |
| 6 | 1991 Oct. 10 | 13 | Custom HCL | Ar | 295 | 15154-36081 | 0.044 | 4 | UV | CuSO$_4$ | S. B. Si PD | Ar I & II, W Strip Lamp |
| 7 | 1991 Oct. 10 | 14 | Custom HCL | Ar | 290 | 15154-36081 | 0.044 | 4 | UV | CuSO$_4$ | S. B. Si PD | Ar I & II, W Strip Lamp |
| 8 | 1991 Oct. 9 | 11 | Custom HCL | Ar | 300 | 7810-25033 | 0.031 | 4 | UV | GG495 | S. B. Si PD | W Strip Lamp |
| 9 | 1991 Oct. 9 | 12 | Custom HCL | Ar | 300 | 7810-25033 | 0.031 | 4 | UV | GG495 | S. B. Si PD | W Strip Lamp |
| 10 | 1991 Dec. 12 | 7 | Custom HCL | Ar | 300 | 1661-11312 | 0.015 | 8 | CaF$_2$ | GaAs | InSb | Ar I & II, W Strip Lamp |
| 11 | 1991 Dec. 12 | 8 | Custom HCL | Ne | 300 | 1661-11312 | 0.015 | 8 | CaF$_2$ | GaAs | InSb | W Strip Lamp |
| 12 | 2001 Jan. 25 | 13 | Commercial HCL | Ne | 25 | 7929-34998 | 0.050 | 10 | UV | | S. B. Si PD | |
| 13 | 2001 Jan. 26 | 15 | Commercial HCL | Ne | 20 | 7929-34998 | 0.050 | 10 | UV | | S. B. Si PD | |
| 14 | 1985 Feb. 6 | 10 | EDL | Ar | | 7456-28808 | 0.035 | 2 | UV | GG375 | S. B. Si PD | W Strip Lamp |
| 15 | 1985 Feb. 6 | 67 | EDL | Ar | | 7456-28808 | 0.035 | 8 | UV | GG375 | S. B. Si PD | W Strip Lamp |

Note.
[a] All spectra were recorded using the 1 m FTS on the McMath telescope at the National Solar Observatory, Kitt Peak, AZ.
[b] Lamp types include commercially available sealed HCLs, a custom water-cooled HCL, and an Electrodeless Discharge Lamp (EDL).
[c] Detector types include the Super Blue (S. B.) Si photodiode (PD).
[d] Relative radiometric calibrations were based on selected sets of Ar I and Ar II lines, on a tungsten-quartz-halogen (WQH) lamp and on a tungsten (W) Strip Lamp.

**Table 2**
3 m Echelle Spectra[a]

| Index[b] | Date | Serial Number | Buffer Gas | Lamp Current (mA) | Frame[c] Desig. | Accum- ulations | Total Exposure (minutes) |
|---|---|---|---|---|---|---|---|
| 21 | 2023 Oct. 30 | 1 | Ar | 10 | B | 95 | 60 |
| 22 | 2023 Oct. 30 | 3 | Ar | 10 | C | 60 | 60 |
| 23 | 2023 Oct. 30 | 5 | Ar | 10 | D | 30 | 60 |
| 24 | 2023 Oct. 30 | 7 | Ar | 10 | C | 60 | 60 |
| 25 | 2023 Nov. 16 | 1 | Ar | 17 | BC | 200 | 50 |
| 26 | 2023 Nov. 16 | 3 | Ar | 17 | C | 200 | 50 |
| 27 | 2023 Nov. 16 | 5 | Ar | 17 | CD | 167 | 50 |
| 28 | 2023 Dec. 6 | 1 | Ar | 20 | B | 90 | 135 |
| 29 | 2023 Dec. 6 | 3 | Ar | 20 | C | 90 | 135 |
| 30 | 2023 Dec. 6 | 5 | Ar | 20 | D | 100 | 150 |
| 31 | 2023 Dec. 6 | 7 | Ar | 20 | B | 90 | 135 |
| 32 | 2023 Dec. 11 | 1 | Ar | 20 | D | 270 | 90 |
| 33 | 2023 Dec. 11 | 3 | Ar | 20 | C | 270 | 90 |
| 34 | 2023 Dec. 11 | 5 | Ar | 20 | D | 180 | 60 |
| 35 | 2023 Dec. 11 | 7 | Ar | 20 | B | 180 | 90 |
| 36 | 2024 Jan. 11 | 1 | Ne | 20 | B | 216 | 90 |
| 37 | 2024 Jan. 11 | 3 | Ne | 20 | C | 180 | 90 |
| 38 | 2024 Jan. 11 | 5 | Ne | 20 | D | 108 | 90 |
| 39 | 2024 Jan. 11 | 7 | Ne | 20 | B | 216 | 90 |
| 40 | 2024 Jan. 22 | 1 | Ne | 25 | C | 160 | 90 |
| 41 | 2024 Jan. 22 | 3 | Ne | 25 | B | 216 | 90 |
| 42 | 2024 Jan. 22 | 5 | Ne | 25 | C | 180 | 90 |
| 43 | 2024 Jan. 22 | 7 | Ne | 25 | D | 180 | 90 |

Notes:

[a] All echelle spectra were taken with commercially manufactured Gd-Ne or Gd-Ar HCLs and have a spectral coverage from 2350 to 4400 Å in the low-resolution direction and resolving power of ~250,000, although the effective resolving power is somewhat lower due to line broadening. Each HCL spectrum was calibrated with a $D_2$ lamp spectrum, which was recorded immediately following the completion of the HCL spectrum.

[b] Each spectrum listed is a single CCD frame and does not cover an entire echelle grating order. A minimum of three overlapping frames are needed per grating order in the UV. Four or five are recorded for each data set to give some redundancy.

[c] Frame Designation of C indicates the CCD frame straddles the center of the grating order, B is shifted toward lower wavelengths and D toward higher wavelengths. The breadth of B-C-D frames covers one grating order in the UV.

Table 3
BFs of Gd II

| Upper Level[a] | | Lower Level[a] | | | | This Study | | DH06[b] | |
|---|---|---|---|---|---|---|---|---|---|
| $E_k$ (cm$^{-1}$) | $J_k$ | $E_i$ (cm$^{-1}$) | $J_i$ | $\lambda_{air}$ Å | $\sigma_{vac}$ (cm$^{-1}$) | BF[c] | Unc. % | BF[c] | Unc. % |
| Even-Parity Upper Levels | | | | | | | | | |
| 30366.818 | 5.5 | 633.273 | 4.5 | 3362.24 | 29733.545 | 0.558 | 1 | 0.559 | 2 |
| $\tau_u$ = 4.2 ± 0.2 ns | 5.5 | 1158.943 | 5.5 | 3422.75 | 29207.875 | 0.0269 | 2 | 0.0307 | 8 |
| | 5.5 | 3972.167 | 4.5 | 3787.57 | 26394.651 | 0.079 | 5 | 0.080 | 5 |
| | 5.5 | 4841.106 | 5.5 | 3916.51 | 25525.712 | 0.141 | 5 | 0.137 | 5 |
| | 5.5 | 4852.304 | 4.5 | 3918.23 | 25514.514 | 0.0083 | 7 | 0.0079 | 6 |
| | 5.5 | 5897.264 | 6.5 | 4085.56 | 24469.554 | 0.143 | 5 | 0.139 | 6 |
| | 5.5 | 8551.049 | 5.5 | 4582.56 | 21815.769 | 0.029 | 9 | 0.029 | 10 |
| | 5.5 | 8884.809 | 4.5 | 4653.75 | 21482.009 | 0.0015 | 14 | 0.0015 | 13 |
| | 5.5 | 11066.865 | 4.5 | 5179.92 | 19299.953 | 0.0041 | 13 | 0.0044 | 19 |
| | 5.5 | 13925.734 | 6.5 | 6080.64 | 16441.084 | 0.008 | 14 | 0.009 | 18 |
| | | | | | residual | 0.000 | | 0.000 | |
| 30996.851 | 6.5 | 1158.943 | 5.5 | 3350.48 | 29837.908 | 0.459 | 2 | 0.459 | 2 |
| $\tau_u$ = 3.7 ± 0.2 ns | 6.5 | 1935.310 | 6.5 | 3439.99 | 29061.541 | 0.244 | 2 | 0.241 | 3 |
| | 6.5 | 4841.106 | 5.5 | 3822.17 | 26155.745 | 0.0074 | 5 | 0.0087 | 10 |
| | 6.5 | 5339.477 | 5.5 | 3896.41 | 25657.374 | 0.0021 | 7 | 0.0027 | 14 |
| | 6.5 | 5897.264 | 6.5 | 3983.00 | 25099.587 | 0.0107 | 8 | 0.0105 | 7 |
| | 6.5 | 6605.154 | 7.5 | 4098.60 | 24391.697 | 0.260 | 4 | 0.248 | 5 |
| | 6.5 | 8551.049 | 5.5 | 4453.93 | 22445.802 | 0.0046 | 11 | 0.0043 | 9 |
| | 6.5 | 11492.204 | 5.5 | 5125.56 | 19504.647 | 0.012 | 15 | 0.012 | 15 |
| | 6.5 | 18753.034 | 7.5 | 8165.14 | 12243.817 | 0.0008 | 24 | 0.0006 | 18 |
| | | | | | residual | 0.000 | | 0.000 | |
| 34178.776 | 5.5 | 633.273 | 4.5 | 2980.16 | 33545.503 | 0.056 | 2 | 0.057 | 7 |
| $\tau_u$ = 2.2 ± 0.2 ns | 5.5 | 1158.943 | 5.5 | 3027.60 | 33019.833 | 0.219 | 1 | 0.211 | 5 |
| | 5.5 | 1935.310 | 6.5 | 3100.50 | 32243.466 | 0.543 | 1 | 0.530 | 1 |
| | 5.5 | 4841.106 | 5.5 | 3407.61 | 29337.670 | 0.069 | 4 | 0.075 | 7 |
| | 5.5 | 5339.477 | 5.5 | 3466.50 | 28839.299 | 0.012 | 9 | 0.015 | 13 |
| | 5.5 | 5897.264 | 6.5 | 3534.87 | 28281.512 | 0.0011 | 20 | ... | ... |
| | 5.5 | 11066.865 | 4.5 | 4325.56 | 23111.911 | 0.033 | 10 | 0.037 | 11 |
| | 5.5 | 11492.204 | 5.5 | 4406.66 | 22686.572 | 0.055 | 10 | 0.055 | 8 |
| | 5.5 | 13377.976 | 5.5 | 4806.16 | 20800.800 | 0.0032 | 16 | 0.0035 | 16 |
| | 5.5 | 13925.734 | 6.5 | 4936.15 | 20253.042 | 0.0065 | 15 | 0.0070 | 20 |
| | 5.5 | 18641.357 | 5.5 | 6434.30 | 15537.419 | 0.0007 | 19 | 0.0006 | 19 |
| | 5.5 | 19401.977 | 4.5 | 6765.50 | 14776.799 | 0.0008 | 20 | 0.0008 | 20 |
| | | | | | residual | 0.000 | | 0.007 | |
| 36845.366 | 2.5 | 3082.011 | 2.5 | 2960.93 | 33763.355 | 0.032 | 7 | ... | ... |
| $\tau_u$ = 4.1 ± 0.2 ns | 2.5 | 3427.274 | 3.5 | 2991.52 | 33418.092 | 0.021 | 8 | ... | ... |
| | 2.5 | 3444.235 | 3.5 | 2993.04 | 33401.131 | 0.044 | 7 | ... | ... |
| | 2.5 | 4483.854 | 3.5 | 3089.19 | 32361.512 | 0.0088 | 11 | ... | ... |
| | 2.5 | 9142.904 | 3.5 | 3608.76 | 27702.462 | 0.436 | 2 | ... | ... |
| | 2.5 | 9328.864 | 2.5 | 3633.15 | 27516.502 | 0.006 | 19 | ... | ... |
| | 2.5 | 9451.697 | 1.5 | 3649.44 | 27393.669 | 0.156 | 3 | ... | ... |
| | 2.5 | 12703.450 | 1.5 | 4141.01 | 24141.916 | 0.041 | 6 | ... | ... |
| | 2.5 | 12776.067 | 2.5 | 4153.50 | 24069.299 | 0.085 | 4 | ... | ... |
| | 2.5 | 12891.692 | 3.5 | 4173.55 | 23953.674 | 0.098 | 6 | ... | ... |
| | 2.5 | 17971.595 | 2.5 | 5296.88 | 18873.771 | 0.007 | 15 | ... | ... |
| | | | | residual | 0.065 | | | | |
| 38029.848 | 4.5 | 3444.235 | 3.5 | 2890.53 | 34585.613 | 0.012 | 12 | ... | ... |
| $\tau_u$ = 7.9 ± 0.4 ns | 4.5 | 4841.106 | 5.5 | 3012.19 | 33188.742 | 0.126 | 5 | ... | ... |
| | 4.5 | 4852.304 | 4.5 | 3013.21 | 33177.544 | 0.010 | 19 | ... | ... |
| | 4.5 | 5339.477 | 5.5 | 3058.12 | 32690.371 | 0.014 | 12 | ... | ... |
| | 4.5 | 8551.049 | 5.5 | 3391.29 | 29478.799 | 0.034 | 20 | ... | ... |
| | 4.5 | 10091.567 | 4.5 | 3578.30 | 27938.281 | 0.060 | 3 | ... | ... |
| | 4.5 | 10391.789 | 3.5 | 3617.17 | 27638.059 | 0.488 | 2 | ... | ... |
| | 4.5 | 13377.976 | 5.5 | 4055.34 | 24651.872 | 0.0081 | 7 | ... | ... |
| | 4.5 | 17725.052 | 5.5 | 4923.57 | 20304.796 | 0.076 | 8 | ... | ... |
| | 4.5 | 18641.357 | 5.5 | 5156.26 | 19388.491 | 0.018 | 13 | ... | ... |
| | 4.5 | 19223.207 | 3.5 | 5315.79 | 18806.641 | 0.023 | 11 | ... | ... |
| | 4.5 | 19401.977 | 4.5 | 5366.81 | 18627.871 | 0.016 | 15 | ... | ... |
| | 4.5 | 19750.111 | 5.5 | 5469.02 | 18279.737 | 0.036 | 16 | | |
| | | | | residual | 0.079 | | | | |
| 38553.210 | 5.5 | 633.273 | 4.5 | 2636.35 | 37919.937 | 0.006 | 18 | ... | ... |
| $\tau_u$ = 4.8 ± 0.2 ns | 5.5 | 4841.106 | 5.5 | 2965.43 | 33712.104 | 0.029 | 7 | ... | ... |
| | 5.5 | 5339.477 | 5.5 | 3009.93 | 33213.733 | 0.006 | 20 | ... | ... |
| | 5.5 | 8551.049 | 5.5 | 3332.13 | 30002.161 | 0.330 | 2 | ... | ... |
| | 5.5 | 8884.809 | 4.5 | 3369.62 | 29668.401 | 0.070 | 3 | ... | ... |
| | 5.5 | 10091.567 | 4.5 | 3512.50 | 28461.643 | 0.376 | 1 | ... | ... |
| | 5.5 | 11066.865 | 4.5 | 3637.13 | 27486.345 | 0.0074 | 7 | ... | ... |
| | 5.5 | 13076.050 | 4.5 | 3923.97 | 25477.160 | 0.0027 | 23 | ... | ... |
| | 5.5 | 13377.976 | 5.5 | 3971.03 | 25175.234 | 0.012 | 20 | ... | ... |
| | 5.5 | 13925.734 | 6.5 | 4059.36 | 24627.476 | 0.068 | 4 | ... | ... |
| | 5.5 | 17725.052 | 5.5 | 4799.85 | 20828.158 | 0.023 | 10 | ... | ... |
| | 5.5 | 18676.965 | 6.5 | 5029.73 | 19876.245 | 0.0039 | 19 | ... | ... |
| | 5.5 | 19401.977 | 4.5 | 5220.14 | 19151.233 | 0.009 | 12 | ... | ... |
| | 5.5 | 19750.111 | 5.5 | 5316.79 | 18803.099 | 0.028 | 11 | ... | ... |
| | | | | residual | 0.029 | | | | |
| 39024.491 | 2.5 | 2856.678 | 1.5 | 2764.07 | 36167.813 | 0.187 | 2 | ... | ... |
| $\tau_u$ = 2.45 ± 0.2 ns | 2.5 | 3082.011 | 2.5 | 2781.40 | 35942.480 | 0.205 | 2 | ... | ... |
| | 2.5 | 3427.274 | 3.5 | 2808.38 | 35597.217 | 0.015 | 7 | ... | ... |
| | 2.5 | 3444.235 | 3.5 | 2809.72 | 35580.256 | 0.388 | 1 | ... | ... |
| | 2.5 | 4027.161 | 1.5 | 2856.52 | 34997.330 | 0.0151 | 4 | ... | ... |
| | 2.5 | 4212.756 | 2.5 | 2871.75 | 34811.735 | 0.0114 | 6 | ... | ... |
| | 2.5 | 4483.854 | 3.5 | 2894.29 | 34540.637 | 0.0049 | 14 | ... | ... |
| | 2.5 | 9328.864 | 2.5 | 3366.53 | 29695.627 | 0.010 | 16 | ... | ... |
| | 2.5 | 9451.697 | 1.5 | 3380.52 | 29572.794 | 0.079 | 7 | ... | ... |
| | 2.5 | 10802.621 | 1.5 | 3542.34 | 28221.870 | 0.028 | 10 | ... | ... |
| | 2.5 | 12776.067 | 2.5 | 3808.67 | 26248.424 | 0.0046 | 12 | ... | ... |
| | 2.5 | 12891.692 | 3.5 | 3825.52 | 26132.799 | 0.0035 | 18 | ... | ... |
| | 2.5 | 17971.595 | 2.5 | 4748.61 | 21052.896 | 0.007 | 24 | ... | ... |
| | 2.5 | 18955.050 | 2.5 | 4981.31 | 20069.441 | 0.010 | 24 | ... | ... |
| | 2.5 | 19223.207 | 3.5 | 5048.77 | 19801.284 | 0.026 | 22 | ... | ... |
| | | | | | residual | 0.006 | | | |
| 39170.192 | 3.5 | 3082.011 | 2.5 | 2770.17 | 36088.181 | 0.079 | 2 | ... | ... |
| $\tau_u$ = 2.55 ± 0.2 ns | 3.5 | 3427.274 | 3.5 | 2796.93 | 35742.918 | 0.426 | 1 | ... | ... |
| | 3.5 | 3444.235 | 3.5 | 2798.26 | 35725.957 | 0.0122 | 5 | ... | ... |
| | 3.5 | 3972.167 | 4.5 | 2840.23 | 35198.025 | 0.201 | 1 | ... | ... |
| | 3.5 | 4212.756 | 2.5 | 2859.78 | 34957.436 | 0.0076 | 6 | ... | ... |
| | 3.5 | 4483.854 | 3.5 | 2882.13 | 34686.338 | 0.0129 | 6 | ... | ... |
| | 3.5 | 4852.304 | 4.5 | 2913.08 | 34317.888 | 0.0127 | 6 | ... | ... |
| | 3.5 | 8884.809 | 4.5 | 3300.97 | 30285.383 | 0.0047 | 14 | ... | ... |
| | 3.5 | 9142.904 | 3.5 | 3329.35 | 30027.288 | 0.054 | 7 | ... | ... |
| | 3.5 | 9328.864 | 2.5 | 3350.09 | 29841.328 | 0.077 | 6 | ... | ... |
| | 3.5 | 10633.083 | 2.5 | 3503.21 | 28537.109 | 0.039 | 12 | ... | ... |
| | 3.5 | 13076.050 | 4.5 | 3831.19 | 26094.142 | 0.007 | 15 | ... | ... |
| | 3.5 | 17817.123 | 4.5 | 4681.86 | 21353.069 | 0.011 | 20 | ... | ... |
| | 3.5 | 17971.595 | 2.5 | 4715.97 | 21198.597 | 0.007 | 19 | ... | ... |
| | 3.5 | 18150.637 | 3.5 | 4756.14 | 21019.555 | 0.0031 | 26 | ... | ... |
| | 3.5 | 18955.050 | 2.5 | 4945.41 | 20215.142 | 0.007 | 19 | ... | ... |
| | 3.5 | 19401.977 | 4.5 | 5057.22 | 19768.215 | 0.012 | 19 | ... | ... |
| | 3.5 | 19946.775 | 4.5 | 5200.54 | 19223.417 | 0.014 | 20 | ... | ... |
| | | | | | residual | 0.013 | | | |
| 39250.737 | 4.5 | 3427.274 | 3.5 | 2790.64 | 35823.463 | 0.0054 | 9 | ... | ... |
| $\tau_u$ = 3.2 ± 0.2 ns | 4.5 | 3444.235 | 3.5 | 2791.97 | 35806.502 | 0.101 | 3 | ... | ... |
| | 4.5 | 3972.167 | 4.5 | 2833.75 | 35278.570 | 0.092 | 2 | ... | ... |
| | 4.5 | 4841.106 | 5.5 | 2905.31 | 34409.631 | 0.207 | 2 | ... | ... |
| | 4.5 | 4852.304 | 4.5 | 2906.26 | 34398.433 | 0.0084 | 5 | ... | ... |
| | 4.5 | 5339.477 | 5.5 | 2948.01 | 33911.260 | 0.0246 | 3 | ... | ... |
| | 4.5 | 8551.049 | 5.5 | 3256.42 | 30699.688 | 0.012 | 10 | ... | ... |
| | 4.5 | 8884.809 | 4.5 | 3292.22 | 30365.928 | 0.178 | 4 | ... | ... |
| | 4.5 | 9142.904 | 3.5 | 3320.44 | 30107.833 | 0.059 | 5 | ... | ... |
| | 4.5 | 10091.567 | 4.5 | 3428.47 | 29159.170 | 0.114 | 6 | ... | ... |
| | 4.5 | 10391.789 | 3.5 | 3464.14 | 28858.948 | 0.081 | 15 | ... | ... |
| | 4.5 | 10599.743 | 3.5 | 3489.28 | 28650.994 | 0.0059 | 8 | ... | ... |
| | 4.5 | 11066.865 | 4.5 | 3547.11 | 28183.872 | 0.0019 | 24 | ... | ... |
| | 4.5 | 13076.050 | 4.5 | 3819.40 | 26174.687 | 0.006 | 17 | ... | ... |
| | 4.5 | 13377.976 | 5.5 | 3863.97 | 25872.761 | 0.016 | 11 | ... | ... |
| | 4.5 | 17817.123 | 4.5 | 4664.26 | 21433.614 | 0.029 | 15 | ... | ... |
| | 4.5 | 17869.878 | 3.5 | 4675.77 | 21380.859 | 0.010 | 15 | ... | ... |
| | 4.5 | 18319.239 | 4.5 | 4776.15 | 20931.498 | 0.006 | 19 | ... | ... |
| | 4.5 | 19223.207 | 3.5 | 4991.73 | 20027.530 | 0.0032 | 24 | ... | ... |
| | 4.5 | 19401.977 | 4.5 | 5036.69 | 19848.760 | 0.005 | 23 | ... | ... |
| | 4.5 | 20047.344 | 3.5 | 5205.96 | 19203.393 | 0.006 | 22 | ... | ... |
| | | | | | residual | 0.029 | | | |
| 39537.159 | 4.5 | 3427.274 | 3.5 | 2768.51 | 36109.885 | 0.018 | 8 | ... | ... |
| $\tau_u$ = 3.2 ± 0.2 ns | 4.5 | 3444.235 | 3.5 | 2769.81 | 36092.924 | 0.096 | 17 | ... | ... |
| | 4.5 | 3972.167 | 4.5 | 2810.93 | 35564.992 | 0.064 | 6 | ... | ... |
| | 4.5 | 4841.106 | 5.5 | 2881.33 | 34696.053 | 0.207 | 5 | ... | ... |
| | 4.5 | 4852.304 | 4.5 | 2882.26 | 34684.855 | 0.004 | 14 | ... | ... |
| | 4.5 | 5339.477 | 5.5 | 2923.32 | 34197.682 | 0.027 | 6 | ... | ... |
| | 4.5 | 8551.049 | 5.5 | 3226.32 | 30986.110 | 0.076 | 4 | ... | ... |
| | 4.5 | 9142.904 | 3.5 | 3289.15 | 30394.255 | 0.0125 | 6 | ... | ... |
| | 4.5 | 10091.567 | 4.5 | 3395.12 | 29445.592 | 0.348 | 2 | ... | ... |
| | 4.5 | 17725.052 | 5.5 | 4583.33 | 21812.107 | 0.023 | 13 | ... | ... |
| | 4.5 | 18150.637 | 3.5 | 4674.53 | 21386.522 | 0.009 | 13 | ... | ... |
| | 4.5 | 18319.239 | 4.5 | 4711.68 | 21217.920 | 0.015 | 13 | ... | ... |
| | 4.5 | 19223.207 | 3.5 | 4921.35 | 20313.952 | 0.0048 | 19 | ... | ... |
| | 4.5 | 19401.977 | 4.5 | 4965.05 | 20135.182 | 0.025 | 14 | ... | ... |
| | 4.5 | 19750.111 | 5.5 | 5052.40 | 19787.048 | 0.031 | 13 | ... | ... |
| | 4.5 | 24113.296 | 3.5 | 6481.67 | 15423.863 | 0.006 | 18 | ... | ... |
| | | | | | residual | 0.034 | | | |
| 40773.207 | 6.5 | 1935.310 | 6.5 | 2574.03 | 38837.897 | 0.0023 | 19 | ... | ... |
| $\tau_u$ = 4.3 ± 0.2 ns | 6.5 | 4841.106 | 5.5 | 2782.21 | 35932.101 | 0.0055 | 7 | ... | ... |
| | 6.5 | 8551.049 | 5.5 | 3102.55 | 32222.158 | 0.543 | 3 | ... | ... |
| | 6.5 | 11492.204 | 5.5 | 3414.20 | 29281.003 | 0.032 | 5 | ... | ... |
| | 6.5 | 13377.976 | 5.5 | 3649.23 | 27395.231 | 0.026 | 7 | ... | ... |
| | 6.5 | 13925.734 | 6.5 | 3723.69 | 26847.473 | 0.057 | 7 | ... | ... |
| | 6.5 | 17725.052 | 5.5 | 4337.52 | 23048.155 | 0.025 | 23 | ... | ... |
| | 6.5 | 18641.357 | 5.5 | 4517.11 | 22131.850 | 0.027 | 12 | ... | ... |
| | 6.5 | 18676.965 | 6.5 | 4524.39 | 22096.242 | 0.024 | 12 | ... | ... |
| | 6.5 | 18753.034 | 7.5 | 4540.02 | 22020.173 | 0.233 | 9 | ... | ... |
| | 6.5 | 19750.111 | 5.5 | 4755.34 | 21023.096 | 0.020 | 14 | ... | ... |
| | | | | | residual | 0.005 | | | |
| | | | | Odd-Parity Level | | | | | |
| 38010.603 | 5.5 | 7992.268 | 6.5 | 3330.34 | 30018.335 | 0.520 | 2 | ... | ... |
| $\tau_u$ = 13.7 ± 0.2 ns | 5.5 | 9092.491 | 5.5 | 3457.05 | 28918.112 | 0.226 | 5 | ... | ... |
| | 5.5 | 9943.779 | 5.5 | 3561.91 | 28066.824 | 0.031 | 6 | ... | ... |
| | 5.5 | 11343.525 | 4.5 | 3748.88 | 26667.078 | 0.152 | 4 | ... | ... |
| | 5.5 | 20093.245 | 5.5 | 5579.63 | 17917.358 | 0.029 | 21 | ... | ... |
| | 5.5 | 21157.496 | 6.5 | 5931.98 | 16853.107 | 0.018 | 26 | ... | ... |
| | | | | | residual | 0.024 | | | |

[a] Level energy and *J* values are taken from NIST ASD. The levels are ordered by parity and increasing energy. Ritz wavelengths are calculated from the energy levels using the five parameter formula for the standard index of air from Peck & Reeder (1972).

[b] DH06: Den Hartog et al. (2006), BFs are calculated from their *A*-values. Uncertainties are those quoted for their *A*-values reduced in quadrature by the lifetime uncertainty.

[c] BFs are given to three figures past the decimal except for cases where the absolute uncertainty is less than 0.001, in which case it is given to four places past the decimal. Occasionally in this table the BFs and residuals do not quite add to 1. This is due to rounding errors, but this discrepancy is well within the uncertainties.

[d] This transition was blended with another line in our ECH spectra. The blend is with a line of Gd I or Gd II. A least-squares analysis was used to determine the blend fraction in each spectrum. See text for further discussion.

Table 4.
*A*-values and log(*gf*)s for 156 transitions of Gd II

| $\lambda_{air}$[a] (Å) | $E_k$[b] (cm$^{-1}$) | $J_k$ | $E_i$ (cm$^{-1}$) | $J_i$ | $A_{ki}$ (10$^6$ s$^{-1}$) | $A_{ki}$ Unc. (10$^6$ s$^{-1}$) | log(*gf*) |
|---|---|---|---|---|---|---|---|
| 2574.03 | 40773.207 | 6.5 | 1935.310 | 6.5 | 0.53 | 0.11 | -2.13 |
| 2636.35 | 38553.210 | 5.5 | 633.273 | 4.5 | 1.19 | 0.23 | -1.83 |
| 2764.07 | 39024.491 | 2.5 | 2856.678 | 1.5 | 76 | 6 | -0.28 |
| 2768.51 | 39537.159 | 4.5 | 3427.274 | 3.5 | 5.6 | 0.6 | -1.19 |
| 2769.81 | 39537.159 | 4.5 | 3444.235 | 3.5 | 30 | 3 | -0.46 |
| 2770.17 | 39170.192 | 3.5 | 3082.011 | 2.5 | 31 | 3 | -0.55 |
| 2781.40 | 39024.491 | 2.5 | 3082.011 | 2.5 | 83 | 7 | -0.24 |
| 2782.21 | 40773.207 | 6.5 | 4841.106 | 5.5 | 1.28 | 0.12 | -1.68 |
| 2790.64 | 39250.737 | 4.5 | 3427.274 | 3.5 | 1.68 | 0.19 | -1.71 |
| 2791.97 | 39250.737 | 4.5 | 3444.235 | 3.5 | 31.5 | 2.1 | -0.43 |
| 2796.93 | 39170.192 | 3.5 | 3427.274 | 3.5 | 167 | 13 | 0.20 |
| 2798.26 | 39170.192 | 3.5 | 3444.235 | 3.5 | 4.8 | 0.5 | -1.35 |
| 2808.38 | 39024.491 | 2.5 | 3427.274 | 3.5 | 5.9 | 0.6 | -1.38 |
| 2809.72 | 39024.491 | 2.5 | 3444.235 | 3.5 | 158 | 13 | 0.05 |
| 2810.93 | 39537.159 | 4.5 | 3972.167 | 4.5 | 19.9 | 1.8 | -0.63 |
| 2833.75 | 39250.737 | 4.5 | 3972.167 | 4.5 | 28.6 | 1.9 | -0.46 |
| 2840.23 | 39170.192 | 3.5 | 3972.167 | 4.5 | 79 | 6 | -0.12 |
| 2856.52 | 39024.491 | 2.5 | 4027.161 | 1.5 | 6.2 | 0.6 | -1.34 |
| 2859.78 | 39170.192 | 3.5 | 4212.756 | 2.5 | 3.0 | 0.3 | -1.53 |
| 2871.75 | 39024.491 | 2.5 | 4212.756 | 2.5 | 4.7 | 0.5 | -1.46 |
| 2881.33 | 39537.159 | 4.5 | 4841.106 | 5.5 | 65 | 5 | -0.09 |
| 2882.13 | 39170.192 | 3.5 | 4483.854 | 3.5 | 5.1 | 0.5 | -1.30 |
| 2882.26 | 39537.159 | 4.5 | 4852.304 | 4.5 | 1.24 | 0.19 | -1.81 |
| 2890.53 | 38029.848 | 4.5 | 3444.235 | 3.5 | 1.56 | 0.20 | -1.71 |
| 2894.29 | 39024.491 | 2.5 | 4483.854 | 3.5 | 2.0 | 0.3 | -1.82 |
| 2905.31 | 39250.737 | 4.5 | 4841.106 | 5.5 | 65 | 4 | -0.09 |
| 2906.26 | 39250.737 | 4.5 | 4852.304 | 4.5 | 2.63 | 0.21 | -1.48 |
| 2913.08 | 39170.192 | 3.5 | 4852.304 | 4.5 | 5.0 | 0.5 | -1.30 |
| 2923.32 | 39537.159 | 4.5 | 5339.477 | 5.5 | 8.3 | 0.7 | -0.97 |
| 2948.01 | 39250.737 | 4.5 | 5339.477 | 5.5 | 7.7 | 0.5 | -1.00 |
| 2960.93 | 36845.366 | 2.5 | 3082.011 | 2.5 | 7.7 | 0.7 | -1.22 |
| 2965.43 | 38553.210 | 5.5 | 4841.106 | 5.5 | 6.1 | 0.5 | -1.01 |
| 2980.16 | 34178.776 | 5.5 | 633.273 | 4.5 | 25.4 | 2.4 | -0.39 |
| 2991.52 | 36845.366 | 2.5 | 3427.274 | 3.5 | 5.1 | 0.5 | -1.38 |
| 2993.04 | 36845.366 | 2.5 | 3444.235 | 3.5 | 10.6 | 0.9 | -1.07 |
| 3009.93 | 38553.210 | 5.5 | 5339.477 | 5.5 | 1.35 | 0.18 | -1.66 |
| 3012.19 | 38029.848 | 4.5 | 4841.106 | 5.5 | 16.0 | 1.2 | -0.66 |
| 3013.21 | 38029.848 | 4.5 | 4852.304 | 4.5 | 1.23 | 0.25 | -1.77 |
| 3027.60 | 34178.776 | 5.5 | 1158.943 | 5.5 | 100 | 9 | 0.22 |
| 3058.12 | 38029.848 | 4.5 | 5339.477 | 5.5 | 1.80 | 0.23 | -1.60 |
| 3089.19 | 36845.366 | 2.5 | 4483.854 | 3.5 | 2.1 | 0.3 | -1.73 |
| 3100.50 | 34178.776 | 5.5 | 1935.310 | 6.5 | 247 | 23 | 0.63 |
| 3102.55 | 40773.207 | 6.5 | 8551.049 | 5.5 | 126 | 7 | 0.41 |
| 3226.32 | 39537.159 | 4.5 | 8551.049 | 5.5 | 23.9 | 1.8 | -0.43 |
| 3256.42 | 39250.737 | 4.5 | 8551.049 | 5.5 | 3.8 | 0.4 | -1.23 |
| 3289.15 | 39537.159 | 4.5 | 9142.904 | 3.5 | 3.9 | 0.3 | -1.20 |
| 3292.22 | 39250.737 | 4.5 | 8884.809 | 4.5 | 56 | 4 | -0.04 |
| 3300.97 | 39170.192 | 3.5 | 8884.809 | 4.5 | 1.8 | 0.3 | -1.62 |
| 3320.44 | 39250.737 | 4.5 | 9142.904 | 3.5 | 18.5 | 1.5 | -0.51 |
| 3329.35 | 39170.192 | 3.5 | 9142.904 | 3.5 | 21.2 | 2.2 | -0.55 |
| 3330.34 | 38010.603 | 5.5 | 7992.268 | 6.5 | 37.9 | 2.0 | -0.12 |
| 3332.13 | 38553.210 | 5.5 | 8551.049 | 5.5 | 69 | 4 | 0.14 |
| 3350.09 | 39170.192 | 3.5 | 9328.864 | 2.5 | 30 | 3 | -0.39 |
| 3350.48 | 30996.851 | 6.5 | 1158.943 | 5.5 | 124 | 7 | 0.47 |
| 3362.24 | 30366.818 | 5.5 | 633.273 | 4.5 | 133 | 7 | 0.43 |
| 3366.53 | 39024.491 | 2.5 | 9328.864 | 2.5 | 4.2 | 0.8 | -1.37 |
| 3369.62 | 38553.210 | 5.5 | 8884.809 | 4.5 | 14.5 | 0.8 | -0.53 |
| 3380.52 | 39024.491 | 2.5 | 9451.697 | 1.5 | 32 | 3 | -0.48 |
| 3391.29 | 38029.848 | 4.5 | 8551.049 | 5.5 | 4.3 | 0.3 | -1.13 |
| 3395.12 | 39537.159 | 4.5 | 10091.567 | 4.5 | 109 | 7 | 0.27 |
| 3407.61 | 34178.776 | 5.5 | 4841.106 | 5.5 | 31 | 3 | -0.18 |
| 3414.20 | 40773.207 | 6.5 | 11492.204 | 5.5 | 7.6 | 0.6 | -0.73 |
| 3422.75 | 30366.818 | 5.5 | 1158.943 | 5.5 | 6.4 | 0.3 | -0.87 |
| 3428.47 | 39250.737 | 4.5 | 10091.567 | 4.5 | 36 | 3 | -0.20 |
| 3439.99 | 30996.851 | 6.5 | 1935.310 | 6.5 | 66 | 4 | 0.21 |
| 3457.05 | 38010.603 | 5.5 | 9092.491 | 5.5 | 16.5 | 1.1 | -0.45 |
| 3464.14 | 39250.737 | 4.5 | 10391.789 | 3.5 | 25 | 4 | -0.34 |
| 3466.50 | 34178.776 | 5.5 | 5339.477 | 5.5 | 5.4 | 0.7 | -0.94 |
| 3489.28 | 39250.737 | 4.5 | 10599.743 | 3.5 | 1.84 | 0.19 | -1.47 |
| 3503.21 | 39170.192 | 3.5 | 10633.083 | 2.5 | 15.3 | 2.2 | -0.65 |
| 3512.50 | 38553.210 | 5.5 | 10091.567 | 4.5 | 78 | 4 | 0.24 |
| 3534.87 | 34178.776 | 5.5 | 5897.264 | 6.5 | 0.50 | 0.11 | -1.95 |
| 3542.34 | 39024.491 | 2.5 | 10802.621 | 1.5 | 11.4 | 1.5 | -0.89 |
| 3547.11 | 39250.737 | 4.5 | 11066.865 | 4.5 | 0.60 | 0.15 | -1.95 |
| 3561.91 | 38010.603 | 5.5 | 9943.779 | 5.5 | 2.27 | 0.18 | -1.28 |
| 3578.30 | 38029.848 | 4.5 | 10091.567 | 4.5 | 7.7 | 0.5 | -0.83 |
| 3608.76 | 36845.366 | 2.5 | 9142.904 | 3.5 | 106 | 6 | 0.10 |

| | | | | | | |
|---|---|---|---|---|---|---|
| 3617.17 | 38029.848 | 4.5 | 10391.789 | 3.5 | 62 | 3 | 0.08 |
| 3633.15 | 36845.366 | 2.5 | 9328.864 | 2.5 | 1.4 | 0.3 | -1.77 |
| 3637.13 | 38553.210 | 5.5 | 11066.865 | 4.5 | 1.54 | 0.14 | -1.44 |
| 3649.23 | 40773.207 | 6.5 | 13377.976 | 5.5 | 6.0 | 0.5 | -0.77 |
| 3649.44 | 36845.366 | 2.5 | 9451.697 | 1.5 | 38.1 | 2.3 | -0.34 |
| 3723.69 | 40773.207 | 6.5 | 13925.734 | 6.5 | 13.3 | 1.2 | -0.41 |
| 3748.88 | 38010.603 | 5.5 | 11343.525 | 4.5 | 11.1 | 0.7 | -0.55 |
| 3787.57 | 30366.818 | 5.5 | 3972.167 | 4.5 | 18.9 | 1.3 | -0.31 |
| 3808.67 | 39024.491 | 2.5 | 12776.067 | 2.5 | 1.9 | 0.3 | -1.61 |
| 3819.40 | 39250.737 | 4.5 | 13076.050 | 4.5 | 1.9 | 0.3 | -1.39 |
| 3822.17 | 30996.851 | 6.5 | 4841.106 | 5.5 | 2.01 | 0.15 | -1.21 |
| 3825.52 | 39024.491 | 2.5 | 12891.692 | 3.5 | 1.4 | 0.3 | -1.73 |
| 3831.19 | 39170.192 | 3.5 | 13076.050 | 4.5 | 2.9 | 0.5 | -1.30 |
| 3863.97 | 39250.737 | 4.5 | 13377.976 | 5.5 | 4.9 | 0.6 | -0.96 |
| 3896.41 | 30996.851 | 6.5 | 5339.477 | 5.5 | 0.56 | 0.05 | -1.75 |
| 3916.51 | 30366.818 | 5.5 | 4841.106 | 5.5 | 33.7 | 2.4 | -0.03 |
| 3918.23 | 30366.818 | 5.5 | 4852.304 | 4.5 | 1.97 | 0.17 | -1.26 |
| 3923.97 | 38553.210 | 5.5 | 13076.050 | 4.5 | 0.57 | 0.13 | -1.80 |
| 3971.03 | 38553.210 | 5.5 | 13377.976 | 5.5 | 2.4 | 0.3 | -1.17 |
| 3983.00 | 30996.851 | 6.5 | 5897.264 | 6.5 | 2.9 | 0.3 | -1.02 |
| 4055.34 | 38029.848 | 4.5 | 13377.976 | 5.5 | 1.02 | 0.09 | -1.60 |
| 4059.36 | 38553.210 | 5.5 | 13925.734 | 6.5 | 14.1 | 0.9 | -0.38 |
| 4085.56 | 30366.818 | 5.5 | 5897.264 | 6.5 | 34.2 | 2.4 | 0.01 |
| 4098.60 | 30996.851 | 6.5 | 6605.154 | 7.5 | 70 | 5 | 0.39 |
| 4141.01 | 36845.366 | 2.5 | 12703.450 | 1.5 | 10.0 | 0.8 | -0.81 |
| 4153.50 | 36845.366 | 2.5 | 12776.067 | 2.5 | 20.8 | 1.4 | -0.49 |
| 4173.55 | 36845.366 | 2.5 | 12891.692 | 3.5 | 23.8 | 1.9 | -0.43 |
| 4325.56 | 34178.776 | 5.5 | 11066.865 | 4.5 | 15.2 | 2.0 | -0.29 |
| 4337.52 | 40773.207 | 6.5 | 17725.052 | 5.5 | 5.8 | 0.8 | -0.64 |
| 4406.66 | 34178.776 | 5.5 | 11492.204 | 5.5 | 25 | 3 | -0.06 |
| 4453.93 | 30996.851 | 6.5 | 8551.049 | 5.5 | 1.24 | 0.15 | -1.29 |
| 4517.11 | 40773.207 | 6.5 | 18641.357 | 5.5 | 6.3 | 0.8 | -0.57 |
| 4524.39 | 40773.207 | 6.5 | 18676.965 | 6.5 | 5.6 | 0.7 | -0.62 |
| 4540.02 | 40773.207 | 6.5 | 18753.034 | 7.5 | 54 | 5 | 0.37 |
| 4582.56 | 30366.818 | 5.5 | 8551.049 | 5.5 | 7.0 | 0.7 | -0.58 |
| 4583.33 | 39537.159 | 4.5 | 17725.052 | 5.5 | 7.1 | 1.1 | -0.65 |
| 4653.75 | 30366.818 | 5.5 | 8884.809 | 4.5 | 0.36 | 0.05 | -1.86 |
| 4664.26 | 39250.737 | 4.5 | 17817.123 | 4.5 | 8.9 | 1.5 | -0.53 |
| 4674.53 | 39537.159 | 4.5 | 18150.637 | 3.5 | 2.7 | 0.4 | -1.05 |
| 4675.77 | 39250.737 | 4.5 | 17869.878 | 3.5 | 3.2 | 0.5 | -0.98 |
| 4681.86 | 39170.192 | 3.5 | 17817.123 | 4.5 | 4.2 | 0.9 | -0.95 |
| 4711.68 | 39537.159 | 4.5 | 18319.239 | 4.5 | 4.8 | 0.7 | -0.80 |
| 4715.97 | 39170.192 | 3.5 | 17971.595 | 2.5 | 2.9 | 0.6 | -1.11 |
| 4748.61 | 39024.491 | 2.5 | 17971.595 | 2.5 | 2.8 | 0.7 | -1.25 |
| 4755.34 | 40773.207 | 6.5 | 19750.111 | 5.5 | 4.8 | 0.7 | -0.65 |
| 4756.14 | 39170.192 | 3.5 | 18150.637 | 3.5 | 1.2 | 0.3 | -1.49 |
| 4776.15 | 39250.737 | 4.5 | 18319.239 | 4.5 | 1.8 | 0.4 | -1.22 |
| 4799.85 | 38553.210 | 5.5 | 17725.052 | 5.5 | 4.8 | 0.6 | -0.70 |
| 4806.16 | 34178.776 | 5.5 | 13377.976 | 5.5 | 1.5 | 0.3 | -1.21 |
| 4921.35 | 39537.159 | 4.5 | 19223.207 | 3.5 | 1.5 | 0.3 | -1.27 |
| 4923.57 | 38029.848 | 4.5 | 17725.052 | 5.5 | 9.6 | 0.9 | -0.46 |
| 4936.15 | 34178.776 | 5.5 | 13925.734 | 6.5 | 2.9 | 0.5 | -0.89 |
| 4945.41 | 39170.192 | 3.5 | 18955.050 | 2.5 | 2.7 | 0.6 | -1.11 |
| 4965.05 | 39537.159 | 4.5 | 19401.977 | 4.5 | 7.9 | 1.2 | -0.53 |
| 4981.31 | 39024.491 | 2.5 | 18955.050 | 2.5 | 4.0 | 1.0 | -1.05 |
| 4991.73 | 39250.737 | 4.5 | 19223.207 | 3.5 | 1.0 | 0.3 | -1.42 |
| 5029.73 | 38553.210 | 5.5 | 18676.965 | 6.5 | 0.81 | 0.16 | -1.44 |
| 5036.69 | 39250.737 | 4.5 | 19401.977 | 4.5 | 1.5 | 0.3 | -1.26 |
| 5048.77 | 39024.491 | 2.5 | 19223.207 | 3.5 | 10.5 | 2.5 | -0.62 |
| 5052.40 | 39537.159 | 4.5 | 19750.111 | 5.5 | 9.6 | 1.4 | -0.44 |
| 5057.22 | 39170.192 | 3.5 | 19401.977 | 4.5 | 4.6 | 1.0 | -0.85 |
| 5125.56 | 30996.851 | 6.5 | 11492.204 | 5.5 | 3.2 | 0.5 | -0.75 |
| 5156.26 | 38029.848 | 4.5 | 18641.357 | 5.5 | 2.3 | 0.3 | -1.04 |
| 5179.92 | 30366.818 | 5.5 | 11066.865 | 4.5 | 0.98 | 0.13 | -1.33 |
| 5200.54 | 39170.192 | 3.5 | 19946.775 | 4.5 | 5.4 | 1.2 | -0.75 |
| 5205.96 | 39250.737 | 4.5 | 20047.344 | 3.5 | 2.0 | 0.5 | -1.09 |
| 5220.14 | 38553.210 | 5.5 | 19401.977 | 4.5 | 2.0 | 0.3 | -1.01 |
| 5296.88 | 36845.366 | 2.5 | 17971.595 | 2.5 | 1.7 | 0.3 | -1.37 |
| 5315.79 | 38029.848 | 4.5 | 19223.207 | 3.5 | 2.9 | 0.3 | -0.92 |
| 5316.79 | 38553.210 | 5.5 | 19750.111 | 5.5 | 5.8 | 0.7 | -0.53 |
| 5366.81 | 38029.848 | 4.5 | 19401.977 | 4.5 | 2.0 | 0.3 | -1.07 |
| 5469.02 | 38029.848 | 4.5 | 19750.111 | 5.5 | 4.6 | 0.8 | -0.69 |
| 5579.63 | 38010.603 | 5.5 | 20093.245 | 5.5 | 2.1 | 0.4 | -0.93 |
| 5931.98 | 38010.603 | 5.5 | 21157.496 | 6.5 | 1.3 | 0.3 | -1.07 |
| 6080.64 | 30366.818 | 5.5 | 13925.734 | 6.5 | 2.0 | 0.3 | -0.88 |
| 6434.30 | 34178.776 | 5.5 | 18641.357 | 5.5 | 0.32 | 0.07 | -1.62 |
| 6481.67 | 39537.159 | 4.5 | 24113.296 | 3.5 | 2.0 | 0.4 | -0.91 |
| 6765.50 | 34178.776 | 5.5 | 19401.977 | 4.5 | 0.37 | 0.08 | -1.51 |
| 8165.14 | 30996.851 | 6.5 | 18753.034 | 7.5 | 0.21 | 0.05 | -1.53 |

Notes: [a] Ritz wavelengths in air determined from the energy levels using the five-parameter formula for the standard index of air taken from Peck & Reeder (1972).
[b] Energy levels and $J$-values from the NIST ASD (Kramida et al. 2024).

Table 5
Gd Abundances in Two Metal-Poor Stars

| Wavelength (Å) | E.P. (eV) | log(*gf*) | HD 222925 log ε (Gd)[a] | J2213−5137 log ε (Gd)[a] |
|---|---|---|---|---|
| 2764.07 | 0.35 | -0.28 | 0.99 ± 0.20 | ... |
| 2769.81 | 0.43 | -0.46 | 0.90 ± 0.15 | ... |
| 2770.17 | 0.38 | -0.55 | 0.88 ± 0.15 | ... |
| 2840.23 | 0.49 | -0.12 | 0.95 ± 0.15 | ... |
| 2980.16[b] | 0.08 | -0.39 | 0.78 ± 0.15 | ... |
| 3100.50[b] | 0.24 | 0.63 | 0.79 ± 0.15 | ... |
| 3102.55 | 1.06 | 0.41 | 0.80 ± 0.15 | ... |
| 3292.22 | 1.10 | -0.04 | ... | 1.09 ± 0.20 |
| 3332.14 | 1.06 | 0.14 | ... | 1.09 ± 0.10 |
| 3362.24[b] | 0.08 | 0.43 | ... | 0.94 ± 0.15 |
| 3395.12 | 1.25 | 0.27 | 0.79 ± 0.15 | 1.03 ± 0.10 |
| 3422.75[b] | 0.14 | -0.87 | ... | 1.03 ± 0.15 |
| 3439.99[b] | 0.24 | 0.21 | 0.79 ± 0.15 | 0.97 ± 0.10 |
| 3617.17 | 1.29 | 0.08 | ... | 1.17 ± 0.15 |
| 3787.57[b] | 0.49 | -0.31 | 0.99 ± 0.10 | ... |
| 3916.50[b] | 0.60 | -0.03 | 0.77 ± 0.20 | 1.06 ± 0.10 |
| 4085.56[b] | 0.73 | 0.01 | 0.92 ± 0.10 | 1.03 ± 0.10 |
| 4098.60[b] | 0.82 | 0.39 | ... | 1.04 ± 0.10 |
| 4325.56[b] | 1.37 | -0.29 | ... | 1.11 ± 0.10 |
| 4406.66[b] | 1.42 | -0.06 | ... | 1.08 ± 0.10 |
| 4540.02 | 2.33 | 0.37 | ... | 1.06 ± 0.15 |

Notes:
[a] log ε (Gd) ≡ $\log_{10} (N_{Gd}/N_{H}) + 12.0$
[b] log(*gf*) value was also presented previously in DH06. The log ε abundances in Table 5 reflect the adoption of the new log(*gf*) values from Table 4.